\documentclass[a4paper,UKenglish,cleveref, autoref, thm-restate]{lipics-v2021}

\usepackage{times}
\usepackage{graphicx}
\usepackage{latexsym}
\usepackage{amsmath}
\usepackage{amssymb}
\usepackage{mathtools}
\usepackage{listings}
\usepackage{booktabs}
\usepackage{bookmark}
\usepackage{enumitem}
\usepackage{multicol}
\usepackage{tikz}
\usepackage{tcolorbox}
\usepackage{lipsum}
\usepackage[utf8]{inputenc}
\usepackage{csquotes}
\usepackage[nopostdot,nogroupskip,nonumberlist]{glossaries}
\usepackage{cases}
\usepackage{todonotes}

\DeclareMathOperator{\plor}{;}

\DeclareMathOperator{\plneg}{\setminus +}

\DeclareMathOperator{\aggregate}{\triangleright}

\newcommand{\fragment}[2]{\ensuremath{\mathcal{#1}^{\text{\tiny #2}}}}

\newcommand{\term}{t}
\newcommand{\tree}{G}
\newcommand{\goal}{g}
\newcommand{\constant}[1]{\textsf{#1}}
\newcommand{\json}[1]{\{\!\!\{ #1 \}\!\!\}}
\newacronym{FOL}{FOL}{\emph{first-order logic}}
\newacronym{CWA}{CWA}{\emph{closed-world assumption}}
\newacronym{NRA}{NRA}{\emph{nested relational algebra}}
\newacronym{RA}{RA}{\emph{relational algebra}}
\colorlet{punct}{red!60!black}
\definecolor{lightgrey}{HTML}{EEEEEE}
\definecolor{delim}{RGB}{20,105,176}
\colorlet{numb}{magenta!40!black}
\lstdefinelanguage{json}{
    basicstyle=\normalfont\ttfamily\footnotesize,
    numbers=none,
    showstringspaces=false,
    breaklines=true,
    frame=lines,
    backgroundcolor=\color{lightgrey},
    stringstyle=\color{black}\ttfamily,
    morestring=[b]",
    literate=
     *{0}{{{\color{numb}0}}}{1}
      {1}{{{\color{numb}1}}}{1}
      {2}{{{\color{numb}2}}}{1}
      {3}{{{\color{numb}3}}}{1}
      {4}{{{\color{numb}4}}}{1}
      {5}{{{\color{numb}5}}}{1}
      {6}{{{\color{numb}6}}}{1}
      {7}{{{\color{numb}7}}}{1}
      {8}{{{\color{numb}8}}}{1}
      {9}{{{\color{numb}9}}}{1}
      {:}{{{\color{punct}{:}}}}{1}
      {,}{{{\color{punct}{,}}}}{1}
      {\{}{{{\color{delim}{\{}}}}{1}
      {\}}{{{\color{delim}{\}}}}}{1}
      {[}{{{\color{delim}{[}}}}{1}
      {]}{{{\color{delim}{]}}}}{1},
}

\title{Prolog as a Querying Language for MongoDB}

\author{Daniel Be{\ss}ler}{Institute for Artificial Intelligence, University of Bremen, DE}{danielb@uni-bremen.de}{}{}

\author{Sascha Jongebloed}{Institute for Artificial Intelligence, University of Bremen, DE}{sasjonge@uni-bremen.de}{}{}

\author{Michael Beetz}{Institute for Artificial Intelligence, University of Bremen, DE}{mbeetz@uni-bremen.de}{}{}

\authorrunning{D. Be{\ss}ler and S. Jongebloed and M. Beetz} 

\Copyright{Daniel Be{\ss}ler and Sascha Jongebloed and Michael Beetz} 



\category{} 

\relatedversion{} 

\nolinenumbers 

\EventEditors{John Q. Open and Joan R. Access}
\EventNoEds{2}
\EventLongTitle{42nd Conference on Very Important Topics (CVIT 2016)}
\EventShortTitle{CVIT 2016}
\EventAcronym{CVIT}
\EventYear{2016}
\EventDate{December 24--27, 2016}
\EventLocation{Little Whinging, United Kingdom}
\EventLogo{}
\SeriesVolume{42}
\ArticleNo{23}

\begin{document}
\maketitle

\begin{abstract}

Today's database systems have shown to be capable of supporting AI applications that demand a lot of data processing. To this end, these systems incorporate powerful querying languages that go far beyond the mere retrieval of data, and provide sophisticated facilities for data processing as well. In the case of SQL, the language has been even demonstrated to be Turing-complete in some implementations of the language. In the area of NoSQL databases, a widely adopted one nowadays is the MongoDB database. Queries in MongoDB databases are represented as sequential stages within an aggregation pipeline where each stage defines a transformation of the input data, and passes the transformed data to the next stage. But aggregation queries tend to become rather large for more complex computational problems, lack organization into re-usable pieces, and are thus hard to debug and maintain. We propose a new database querying language called Mongolog which is syntactically a subset of the Prolog language, and we define its operational semantics through translations into aggregation pipelines. To this end, we make use of and extend the formal framework of the MQuery language which characterizes the aggregation framework set-theoretically.

\end{abstract}

\section{Introduction}
\label{sec:intro}

The last decades have seen huge step changes regarding the use of large-scale databases.
An important aspect for the success of systems that use large-scale databases is the scalability of the underlying database to billions of information pieces.
This is often accomplished through the MapReduce programming model~\cite{Dean08}.
It allows to process and generate large data sets in parallel by execution of the mapping function on a cluster of computers.

This raises the question what types of problems can be formulated in a way such that modern databases can efficiently solve them, and how formal frameworks can be integrated with these database systems.
The investigation of this question has been the subject of numerous works that attempt to establish a strong coupling between symbolic AI methods and database technologies.
A well known example is the Datalog language~\cite{Ceri89,Cal2010DatalogAF}
which belongs to the family of logic programming languages.

Logic programming languages exist since several decades, have been investigated thoroughly, and thus have well known properties.
The most widely known language from this family is Prolog, which is defined in an ISO standard~\cite{ISOProlog}.
Prolog is often called \emph{impure} as it cannot be defined within a purely logical framework.
Its semantics is rather defined based on operational characteristics, i.e. based on how it is evaluated according to a resolution strategy~\cite{Vieille87}.


Datalog is syntactically a subset of the Prolog language with additional constraints regarding the use of negation, recursion, and variables appearing in rules.
Several fragments of the language have been investigated such as $\textit{Datalog}^{\neg}$ which is the fragment of Datalog programs with negation~\cite{Ketsman20}.
Datalog has also been extended with additional built-in predicates, and for the use of complex objects~\cite{Ceri89}.
However, some built-in predicates of the Prolog language can only be meaningfully defined with procedural semantics, and, thus, would conflict with the declarative semantics of a Datalog program.
One advantage of purely logical semantics is that logic-based optimization methods, such as \emph{magic sets}~\cite{Balbin88}, can be used to transform a logic program into a more efficient one.
It is further known that Datalog can be translated into relational algebra~\cite{CeriGL86}, and into the SQL language.
However, not all SQL implementations support full Datalog, and are restricted to a subset of the language, e.g., to \emph{linearly recursive} Datalog programs using the recursion construct of the SQL’99 standard.
Datalog is also used as a basis of the \emph{J-logic} framework which was recently proposed to study logical foundations of JSON querying languages~\cite{Hidders20}.
A central notion in the \emph{J-logic} framework are paths which are seen as sequences of keys from which new keys can be constructed through \emph{packing}.

Another recently proposed database querying language based on the logic programming paradigm is \emph{Yedalog}~\cite{Yedalog15}, and its successor language \emph{Logica} which has not been formally specified yet~\footnote{\url{https://logica.dev/}}.
These languages compile to SQL, and can be evaluated by the proprietary \emph{BigQuery} databases
while Mongolog is evaluated via the free, document-based MongoDB.
The motivation behind Yedalog is the coupling of \emph{data-parallel pipelines} and computations in a single language that further makes the handling of semi-structured data more intuitive. 
Yedalog is based on the \emph{Dyna} language~\cite{Eisner11dyna}
which is designed for the specification of data-intense AI systems where intermediate results and conclusions need to be derived from heterogeneous and often uncertain extensional data.
Unlike Yedalog, we strictly consider the well known Prolog ISO standard for the syntax of the language.

The naive evaluation of a Prolog program would sequentially visit all data records that are needed to answer a query given by the user.
But this \emph{one-tuple-at-a-time} fetching of the data is rather slow for non atomic queries due to the input-output processing for each tuple.
However, many database systems allow combining different fetch operations into pipelines that can be optimized and evaluated by the database system.
One of the most widely used database systems nowadays is the MongoDB database.
Its rather powerful querying language is called the \emph{aggregation framework}.
We provide some examples in App. \ref{sec:appendix:mongodb} for the readers who are unfamiliar with the aggregation framework.

MongoDB stores data in form of documents using the JavaScript Object Notation (JSON)~\cite{Bourhis17}.
There is no standard query language for JSON, but the widely adopted aggregation framework of MongoDB has become the de facto standard.
The aggregation framework was designed pragmatically, and has only recently been formalized through the MQuery language~\cite{BotoevaCCX18}.
A query is expressed as a sequence of stages called an \emph{aggregation pipeline} where each stage corresponds to some operator provided by the aggregation framework.
While being certainly expressive, aggregation pipelines also tend to be excessively long, hard to read, and they lack organization of code into reusable chunks -- problems that can be relaxed through the use of Prolog as a database querying language.



In this paper, we investigate a syntactic fragment of the Prolog language, called \emph{Mongolog}, and define its semantics within the formal framework of the MQuery language.
Summarizing, the contributions of this work are the following ones:
\begin{itemize}
    \item an extension of MQuery with additional aggregation operators for the computation of the transitive closure of relations, control over the number of results in an aggregation pipeline, the ordering in which the results are yielded, and an alternative syntax for performing joins (in Section~\ref{sec:mquery});
    
    \item the definition of the Mongolog language as a syntactic fragment of Prolog, and with semantics defined through translations into the MQuery formalism (in Section~\ref{sec:mongolog}).
    These translations are syntactically close to the querying language of the aggregation framework, and, thus, imply an implementation in MongoDB databases; and
    
    \item an investigation of Mongolog-specific optimization techniques for the elimination of predicates and pipeline stages, and the reduction of data passed through the pipeline (in Section~\ref{sec:optimization}).
\end{itemize}


\section{Preliminaries}
\label{sec:preliminaries}

In this section, we introduce notions used throughout this paper.
These are mainly based on the MQuery language ~\cite{BotoevaCCX18} that we use to define semantics of Mongolog queries, and common notions used in the scope of logic programming.

\subsection{The MQuery Language}

\begin{figure}
\centering
\begin{tikzpicture}[auto,
     every node/.style={rectangle,draw,rounded corners=5,anchor=north},
     level/.style={align=center,level distance=8mm},
     edge from parent/.style={draw,-latex},
     mynode/.style={font=\scriptsize\sffamily},
     mylabel/.style={draw=none,swap,anchor=center,font=\scriptsize\ttfamily\bf}
]
  \tikzstyle{level 1}=[sibling distance=30mm]
  \tikzstyle{level 2}=[sibling distance=8mm]
  \tikzstyle{level 3}=[sibling distance=8mm]
\node (1) {$\json{}$} 
    child {node[mynode] (id) {4}  edge from parent node[mylabel] {\_id}}
    child {node[mynode] (refrigator_1) {refrigerator\_1}  edge from parent node[mylabel] {parent}}
    child {node[mynode] (shelf_1) {shelf\_1}  edge from parent node[mylabel] {child}}
        child {node[mynode] (translation) {[]}
            child {node[mynode] {1.5} edge from parent node[mylabel] {0}}
            child {node[mynode] {0.0} edge from parent node[mylabel] {1}}
            child {node[mynode] {3.2} edge from parent node[mylabel] {2}}
            edge from parent node[mylabel] {translation}}
        child {node[mynode] (quaternion) {[]}
            child {node[mynode] {0.0} edge from parent node[mylabel] {0}}
            child {node[mynode] {0.0} edge from parent node[mylabel] {1}}
            child {node[mynode] {0.0} edge from parent node[mylabel] {2}}
            child {node[mynode] {1.0} edge from parent node[mylabel] {3}}
            edge from parent node[mylabel] {quaternion}};
\end{tikzpicture}
\caption{Example tree representing a meronomic relationship between a refrigerator and a shelf.}
\label{fig:tree}
\end{figure}
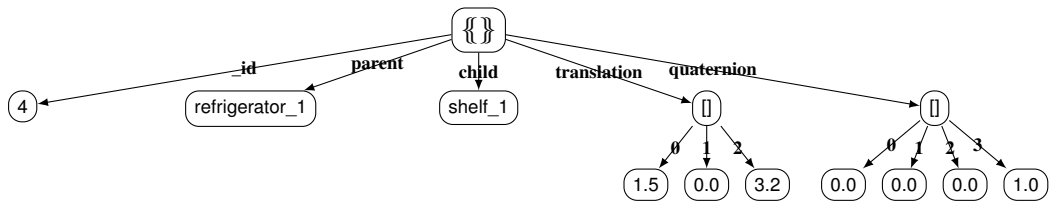

Botoeva et al.~\cite{BotoevaCCX18} have investigated 
formal foundations of the aggregation framework of MongoDB, and its connection to \gls{NRA}.
The authors coin the language of queries that can be formulated within the aggregation framework as \emph{MQuery}, and they investigate the \emph{well-typed} \fragment{M}{MUPGL} fragment of the language that includes the \emph{match}, \emph{unwind}, \emph{project}, \emph{group}, and \emph{lookup} operators.
These operators roughly correspond to \emph{select}, \emph{unnest}, \emph{project}, \emph{nest}, and \emph{left join} operators of NRA respectively.
Within MQuery, a JSON object is seen as a tree of the form $\tree = (N, E, L_n, L_e)$, where $N$ is the set of vertices, and $E$ is the set of edges.
$L_n : N \rightarrow V \cup \{`\json{}`, `[]`\}$ and
$L_e : E \rightarrow K \cup I$ are labeling functions for vertices in $N$ and edges in $E$ respectively,
where
$V$ is a set of \emph{literals}, containing the special elements \constant{null}, \constant{true}, and \constant{false},
$K$ is a set of keys, and
$I$ the set of indices.
Nodes are either labeled by a literal to indicate an atomic value, the constant $\json{}$ to indicate a JSON object, or the constant $[]$ to indicate a JSON array in the subtree underneath the labeled node.
In the following, we will use the notion of JSON objects interchangeably with trees.
The root of $\tree$ is denoted as $\textsf{root}(\tree)$.
An example tree representing spatial data is depicted in Figure~\ref{fig:tree}.
A forest is then characterized set theoretically as a set of trees.

Database collections are seen as forests, and each stage $s$ in an aggregation query $q$ is formalized as an operation $F \aggregate s$ that transforms the input forest $F$.
The whole query is defined as a composition of $n$ pipeline stages:
$q = C \aggregate s_1 \aggregate \dots \aggregate s_n$, where $C$ is the input collection.
Paths are used as parameters of MQuery operators to refer to a sub-tree of the input tree.
Each path is represented as a (possibly empty) concatenation of keys, and
a path $p'$ is a \emph{prefix} of a path $p$ if $p = p'.p''$ for some (non-empty) path $p''$.
The symbol $\epsilon$ is used to denote the empty path.
However, in some cases, the parameter of an operator can be of different forms.
This is captured through the notion of \emph{value definitions}.
A value definition $d$ is represented either as
a constant $v$,
a value reached through a path $p$,
an array with nested value definitions,
a Boolean expression $\beta$, or
a conditional expression $\beta ? d_1 : d_2$ where $d_1,d_2$ are value definitions.
An expression $\beta$ is a Boolean combination of atomic equality conditions between a path and another path, or a constant value, and existential conditions of the form $\exists p$, where $p$ is a path.
A value definition $d$ must be evaluated wrt. a tree $\tree$ in order to obtain the actual parameter value of an operator.
The evaluation, denoted as $\textsf{eval}(d,\tree)$, is defined as:
\begin{eqnarray*}
\begin{aligned}
    &\begin{aligned}[c]
    &d, \text{if $d \in V$};
    \\
    &\textsf{subtree}(\tree, d),
        \text{if $d$ is a path};
    \end{aligned}
    \hspace{0.4cm}
    \begin{aligned}[c]
    &[\textsf{eval}(d_1,\tree), \ldots, \textsf{eval}(d_m, \tree)],
        \text{if $d = [d_1,\ldots,d_m]$};
    \\
    &\text{the value of $\tree \models \beta$},
        \text{if $d=\beta$ is a Boolean value definition}; \text{and}
    \end{aligned}
    \\
    &\textsf{eval}(d',\tree),
        \text{if $d = (\beta ? d_1 : d_2)$, where $d'= d_1$ when $\tree \models \beta$ and $d' = d_2$ otherwise,}
\end{aligned}
\end{eqnarray*}
where $G \models \beta$ iff the Boolean expression $\beta$ is true wrt. tree $G$.

In the following, we rely on definitions in the MQuery language for the different pipeline operators, and auxiliary operators over trees.
We briefly introduce them here, and refer the reader to the work of Botoeva et al. for a formal definition.
The additional MQuery notions used in this paper are:
\begin{itemize}
    \item $\textsf{subtree}(\tree,p)$, which returns the subtree of tree $\tree$ rooted at path $p$ in case the path resolves to a single node.
    The path may resolve to multiple nodes in case it is pointing into an array of objects, in which case the \textsf{subtree} operator yields an array of subtrees instead of a single one.
    E.g., $\textsf{subtree}(\json{\text{a:} [
        \json{\text{b:} 2},
        \json{\text{b:} 3}
    ]}, \text{a}.\text{b})$ returns $[2,3]$;
    
    \item $\textsf{attach}(p,\tree)$, which constructs a new tree that attaches the root of $\tree$ to the end of path $p$.
    E.g., $\textsf{attach}(\text{a},\json{\text{child}: \text{shelf1}})$ returns $\json{\text{a}: \json{\text{child}: \text{shelf1}}}$;
    
    \item $\textsf{array}(F,p)$, which constructs a tree that represents an array of subtrees rooted at path $p$ in trees of forest $F$.
    E.g., $\textsf{array}(\{
        \json{\text{a}: 2},
        \json{\text{a}: 3} \},\text{a})$ constructs the tree
    $\json{\text{1}: 2, \text{2}: 3}$;
    
    \item $t_1 \oplus t_2$, which constructs a tree by merging the trees $t_1$ and $t_2$ based on which nodes are reachable via identical paths.
    E.g., $\json{\text{a}: \json{\text{b}:1}} \oplus \json{\text{a}:\json{\text{c}:2}}$ returns $\json{\text{a}:\json{\text{b}:1, \text{c}:2}}$;
    
    \item \emph{match} $\mu_{\psi}$, which selects a tree if it satisfies the criterion $\psi$ (written as $t \models \psi$).
    $\psi$ is a Boolean combination of atomic conditions of the form
    $d_1 = d_2$,
    $d_1 \subseteq d_2$, and
    $\exists p$, where $d_1,d_2$ are value definitions, and $p$ is a path (see App.~\ref{sec:appendix:match} for details);
    
    \item \emph{unwind} $\omega_{p}$, which \emph{deconstructs} an array reached through path $p$, and yields a tree for each element of the array. The corresponding element of the array replaces the array value at path $p$ in output trees.
    E.g., $\omega_{\text{a}}$ evaluated on $\json{\text{a}: [ 2,3 ] }$ returns the trees $\json{\text{a}: 2 }$ and $\json{\text{a}: 3 }$ (see App.~\ref{sec:appendix:unwind} for details); and
    
    \item \emph{project} $\rho_{P}$, which constructs trees according to the evaluation of sequence $P$ wrt. the input trees.
    Elements of $P$ have the form $p$ or $q/d$, where $p$ is a path to be kept, and $q$ is a new path to be added, or old path to be modified whose value in output documents is defined by $d$.
    E.g., $\rho_{\text{b}/((\exists \text{a}) ? 1 : 2)}$ evaluated on the input tree $\json{\text{a:} 1}$ returns $\json{\text{b:} 1}$ (see App.~\ref{sec:appendix:project} for details).
\end{itemize}

Note that we allow value definitions in equality conditions of the match operator, which is not the case in the original definition in MQuery where only paths may occur on the left side, and constant values on the right side of equality conditions in a match criterion $\varphi$. However, this extension is harmless.
In addition, we introduce a new type of condition written as $d_1 \subseteq d_2$ that succeeds if each value defined by $d_1$ is also defined by $d_2$ in some tree $\tree$, i.e., if $V_1^\tree \subseteq V_2^\tree$, where $V_1^\tree$ is the set of values defined by $d_1$, and $V_2^\tree$ is the set of values defined by $d_2$ in tree $\tree$.

We further introduce a new pipeline operator $\omega^{*}_p$ which is a variant of the \emph{unwind} stage that succeeds even if the path $p$ resolves to the value \constant{null}, or an empty array.
Thus, $\omega_{p}^{*}(\tree) = \tree$ if $\omega_{p}(\tree) = \emptyset$, and $\omega_{p}^{*}(\tree) = \omega_{p}(\tree)$ otherwise, where $\tree$ is the input tree of the operator.

\subsection{Syntax of Prolog Programs}

In the following, we briefly recap the syntax of logic programs as defined in the Prolog ISO standard~\cite{ISOProlog}.
A Prolog program consists of a finite set of \emph{facts} and \emph{rules}.
A rule \emph{"if $b$ is true, then also $h$ is true"} is denoted as $h \leftarrow b$ where $h$ is called \emph{head}, and $b$ \emph{body} of the rule.
The definition of a rule may consist of several such clauses.
The head of the rule $h$ has the form $p(\term_1, \dots, \term_n)$ where $p$ is a \emph{functor} (also called \emph{predicate symbol}) of a $n$-ary predicate, and each $\term_i$ is a \emph{term}.
A term is either a constant value (number or atom), a variable, or a complex structure.
E.g., $p(q(r(x),y))$ is a complex term, where $p,q,r$ are constants (atoms), and $x,y$ are terms.
The body of the rule $b$ consists of a sequence of \emph{goals} separated with a comma: $b = \goal_1, \dots, \goal_m$, where each $\goal_i$ is a non-variable term, i.e., the predicate symbol of a goal must be a constant, and cannot be a variable.
Each predicate symbol either corresponds to a user-defined relation, or has built-in semantics.
Facts are written as rules without body: $p(\term_1, \dots, \term_n) \leftarrow$, where each $\term_i$ is a term.
Prolog programs are then evaluated wrt. a query given by the user.
We denote such a query as $\leftarrow \goal_1, \dots, \goal_n$, where each $\goal_i$ is a goal.

\section{Predicate Literals in JSON Databases}
\label{sec:database}

The evaluation of Mongolog rules is performed wrt. facts that are stored in a MongoDB database.
In this section, we elaborate on how such facts are represented in the database system.
The facts and rules in a logic program $p$ are often stored separately in database oriented logic programming languages.
The storage for ground facts is called \emph{Extensional Database} (EDB).
In the case of Mongolog, the EDB is a document-based storage.
Each document in the EDB represents a fact in $p$.
The different predicates that appear in the EDB are called EDB-predicates.
The set of rules in $p$, on the other hand, is called \emph{Intensional Database} (IDB).
In Mongolog programs, EDB-predicates may not appear in the head of rules,
i.e., a predicate cannot be both an EDB and an IDB predicate.

Predicate symbols of ground facts in the EDB are identified by collection names in the document storage.
Each collection stores the set of facts that share the same predicate symbol and arity in form of key-value documents.
The keys are used to identify the different arguments of the fact.
Let $p(\term_1 \dots \term_k)$ be a ground term, i.e., each $\term_i$ does not contain a variable.
With a JSON-based EDB, this term can be stored in a collection labeled by $p$ as a \emph{term document} of the form $\json{ n_1: c_1, \dots, n_i: c_k }$, where $n_1, \dots, n_i$ identify different arguments of the fact with the associated JSON values $c_1, \dots c_i$.
The collection $C_{p/n}$ that stores $m$ facts of the $n$-ary predicate $p$ can be defined as: $C_{p/n} = \{ D_1, \dots, D_m \}$, where each $D_i$ is a term document.

One advantage of a document-based storage is that nested structures can be represented directly. 
Recursive structures are nevertheless difficult to handle operationally.
For that reason, we employ a flattened syntax of terms where index keys are \emph{packed} into a single key to locate each argument within the nested structure.
E.g., the nested ground term $p(a,q(2))$ is represented in the following way:
$\json{
    \text{0}: \text{"p"},
    \text{1}: \text{"a"},
    \text{2.0}: \text{"q"},
    \text{2.1}: 2
}$.
As a more concrete example, let
us consider a 2-ary parthood relationship stored in a collection $C_\textsf{hasPart/2}$ consisting of the following documents:
\begin{eqnarray*}
\begin{aligned}
    &\json{
        \text{0:}\ \text{"hasPart"},
        \text{1:}\ \text{"fridge1"},
        \text{2:}\ \text{"door1"}
    }, 
    \json{
        \text{0:}\ \text{"hasPart"},
        \text{1:}\ \text{"door1"},
        \text{2:}\ \text{"handle1"}
    }, \text{ and}
    \\
    &\json{
        \text{0:}\ \text{"hasPart"},
        \text{1:}\ \text{"door1"},
        \text{2:}\ \text{"handle2"}
    }.
\end{aligned}
\end{eqnarray*}

The collection $C_\textsf{hasPart/2}$
is a concrete representation of the ground facts $\textsf{hasPart}(\text{fridge1}, \text{door1})$, $\textsf{hasPart}(\text{door1}, \text{handle1})$, and $\textsf{hasPart}(\text{door1}, \text{handle2})$.
The arguments of the predicate represent the parent, and the child of the parthood relationship respectively.
Mongolog rules can then be used to define new relationships based on facts that are stored in collections.
E.g., the rule $\textsf{hasPart}_\textsf{reflexive}(x,y) \leftarrow \textsf{hasPart}(x,y) \plor \textsf{hasPart}(y,x)$ defines a reflexive 2-ary parthood relationship named $\textsf{hasPart}_\textsf{reflexive}$ based on facts stored in the collection $C_\textsf{hasPart/2}$.

Term documents can trivially be translated into array form, where each element in the array consists of an index key and associated value (if any).
We represent the array elements as documents, where the index key of the element is stored at key \emph{k}, and its value at key \emph{v}.
E.g., $p(a)$ has the flattened array form
$[ \json{\text{k}: \text{"0"}, \text{v}: \text{"p"}},
   \json{\text{k}: \text{"1"}, \text{v}: \text{"a"}} ]$.
The array form is useful for certain operations over terms, such as the construction of a sub-term based on index keys (see Section~\ref{sec:mquery} and App.~\ref{sec:appendix:semantics}).

A term that is an argument of a goal in a query may consist of variables.
We use the convention that the value of variables in JSON syntax is the constant \textsf{undefined}, however, an additional field \emph{n} is added whose value is the name of the variable.
E.g., the term $p(x)$, where $x$ is a variable, is represented as the flattened array:
$[ \json{\text{k}: \text{"0"}, \text{v}: \text{"p"}},
   \json{\text{k}: \text{"1"}, \text{n}: \text{"x"}} ]$,
where $\textsf{name}(x) = \text{"x"}$ is the unique name of variable $x$.
Note that the index key of variables would disappear in \emph{term document} form, and, thus, only ground terms may be represented this way.

In the following, we further need an auxiliary function that decomposes terms into their indexed variables.
We denote the set of indexed variables of a term $\term$ as $\textsf{vars}(\term)$. Each element of this set identifies a variable $x$ within $\term$, and has the form $(n,k)$, where $n=\textsf{name}(x)$ is a unique name of $x$, and $k$ its index key in $\term$.
E.g., $\textsf{vars}(p(x))$ returns
$\{ (\text{"x"}, \text{1}) \}$
if $x$ is a variable, and
$\textsf{name}(x) = \text{"x"}$.
The function is formally defined in App.~\ref{sec:appendix:semantics}.

\section{Extensions of the MQuery Language}
\label{sec:mquery}

For our investigation, we take \fragment{M}{MUPG} as a basis, and extend it with the following operators:
\begin{itemize}
    
    \item \emph{sort} $\sigma_{P}$, which yields a forest with a \emph{defined order}. The output forest has exactly the same amount of trees as the input forest. $P$ is a sequence of sort criteria of the form $+p$ or $-p$ that are evaluated in sequence-order, where $p$ is a non-empty path, and the plus and minus signs indicate ascending and descending order respectively. (See App.~\ref{sec:appendix:sort} for details.)
    
    \item \emph{limit} $\kappa_{k}$, which selects $k$ trees from the input forest. In case the forest has a \emph{defined order}, the first $k$ trees are selected according to a sequence of sort criteria, or, in case its order is undefined, arbitrary trees are selected.
    In the latter case, different $k$ trees may be selected in case the \emph{limit} stage is evaluated twice on the same set of input trees. (See App.~\ref{sec:appendix:limit} for details.)
    
    \item \emph{lookup} $\lambda_{p}^{V}[C, s_1 \aggregate \dots \aggregate s_k]$, which runs, for each input tree of the lookup stage, another aggregation pipeline with $k$ stages on trees from collection $C$, and stores the resulting forest as an array at a non-empty path $p$ in the output tree.
    $V$ is a sequence of \emph{variable specifications} of the form $q_i/d_i$, where $q_i$ is a path of a variable whose value within the pipeline is determined by the \emph{value definition} $d_i$, and
    $q_i \neq q_j$ for all $i \neq j$. (See App.~\ref{sec:appendix:lookup} for details.)
    
    \item \emph{graph-lookup} $\tau_{p,p_d,\varphi,D}^{p_1 = p_2}[C]$, which performs a recursive search on collection $C$ for each input tree, and for each of the values defined in the sequence $D = d_1, \dots, d_k$, where each $d_i$ is a value definition.
    Each recursion step selects trees in $C$ that satisfy criterion $\varphi$, and whose path $p_1$ has the given initial value in the first step, or, otherwise, the same value of path $p_2$ in a tree visited in the last step.
    An output tree is produced for each terminal sequence of visited trees, i.e., where the recursion has terminated for the last tree in the sequence.
    The sequence is then stored as an array at a non-empty path $p$ in the output document,
    and a new node indicating the depth of recursion is added to each visited tree
    at path $p_d$. (See App.~\ref{sec:appendix:graphlookup} for details.)
\end{itemize}

The lookup operator considered here is more powerful then the one originally considered in MQuery which only allowed a single equality join condition.
Moreover, we consider the additional operators
\emph{graph-lookup} ($\tau$) for computing the transitive closure of a relation,
\emph{limit} ($\kappa$) providing control over the number of trees that are passed through in the pipeline,
and \emph{sort} ($\sigma$) defining the order of a forest.
We coin the extended fragment of MQuery as $\fragment{M}{MUPGLTKS}_{v3.6}$ where \emph{T} refers to inclusion of transitive closure, \emph{K} to inclusion of the limit operator to select at most $k$ input trees, \emph{S} to the \emph{sort} operator, and $v3.6$ to the (minimum) version of the aggregation framework.

%
The inclusion of the \emph{sort} operator carries the implication that an ordering must be defined for trees in a forest.
But MQuery formalizes pipelines in a set-theoretical way, and, thus, the output of stages cannot be ordered.
Nevertheless, sort criteria can be evaluated when a forest is filtered through the limit operator, and when a forest is presented to the user.
However, sorting in MongoDB is based on values within trees that are processed by the \emph{sort} stage such that ordering is not influenced by projections performed after it, and, therefore, the values must be stored for sorted sampling at a later point.
Let $C=(c_1,\dots,c_n)$ be a sequence of $n>0$ sort criteria.
Then, $n$ values must be stored for each tree by the \emph{sort} stage.
We store these values via an additional node labeling function $L_\sigma: \{ \textsf{root}(\tree) \} \rightarrow V^{*}$ that maps the root node of a tree $\tree$ to a sequence of literals.
Thus, each tree is seen as a tuple $(N,E,L_n,L_e,L_\sigma)$.
The semantics of the \emph{sort} pipeline stage is then that it defines $L_\sigma$ for each of its output trees:
\begin{eqnarray}
F \triangleright \sigma_{C} &=&
    \{ (N, E, L_n, L_e, L_\sigma^{'}) 
    \mid \tree = (N,E,L_n,L_e,L_\sigma) \in F \},
\end{eqnarray}
where $L_\sigma^{'}(\textsf{root}(\tree)) = (\textsf{eval}(c_1,\tree), \dots, \textsf{eval}(c_n,\tree), v_1, \dots, v_k)$,
$L_\sigma(\textsf{root}(\tree))=(v_1,\dots,v_k)$ with $k\geq0$, 
$\textsf{eval}(+p,\tree) = \json{\text{+}: \textsf{eval}(p,\tree)}$, and $\textsf{eval}(-p,\tree) = \json{\text{-}: \textsf{eval}(p,\tree)}$.
We say that a forest has \emph{defined order} if it has a non-empty $\sigma$-label, and \emph{undefined order} otherwise.
Note that the order of database collections is undefined.
Let $\tree_1$ and $\tree_2$ be two trees, and $<_\sigma$ an ordering over trees.
We say that 
$\tree_1 <_\sigma \tree_2$ holds iff $L_\sigma^{1}(\textsf{root}(\tree_i))
    <_\sigma
    L_\sigma^{2}(\textsf{root}(\tree_j))$, where
$L_\sigma^{1}$ and $L_\sigma^{2}$ are the $\sigma$-labeling functions of $\tree_1$ and $\tree_2$ respectively.
We further say that a value sequence
$v_{11}, \dots, v_{1k}$ is smaller then another sequence
$v_{21}, \dots, v_{2n}$ under $<_\sigma$ if there exists $v_{1i}$ with $v_{1i} <_{\mathcal{M}} v_{2i}$, and $v_{1j} =_{\mathcal{M}} v_{2j}$ for all $j < i$, where
$<_{\mathcal{M}}$ denotes the \emph{default order} in MongoDB databases
(see App.~\ref{sec:appendix:ordering}).

The $\sigma$-labeling function has interactions with the \emph{sort}, \emph{limit} and \emph{group} stages of MQuery, but other operators are not influenced by it.
E.g., the $\sigma$-labeling of input trees in a matching stage remains the same for the output trees.
In case of \emph{group} stages, the forest produced has undefined order, i.e., $L_\sigma$ is replaced in each tree with a function that maps the root of the tree to a nullary tuple.
The semantics of the \emph{limit} operator can now be characterized by the set of $n$ \emph{smallest} trees according to the ordering of the input forest:
\begin{eqnarray}
F \triangleright \kappa_{k} &=&
    \{ \tree_1, \dots, \tree_n
    \mid
        \forall 1 \leq i \leq n,
        \forall \tree \in F \setminus \{ \tree_1, \dots, \tree_n \}:
        \tree_i \in F \wedge \tree_i \leq_\sigma \tree
    \},
\end{eqnarray}
where $n$ is the number of trees yielded by the \emph{limit} operator
with $0 \leq n \leq k$,
$n = k$ if $|F| \geq k$, and
$n = |F|$ if $|F| < k$.
Note that subsequent evaluation with the same set of inputs may yield different results in case the ordering is \emph{undefined}, or if there are equal values according to the ordering.

Semantics for the alternate syntax of the \emph{lookup} operator can be defined similarly to the original definition in MQuery.
The forest produced consists exactly of the trees in the input forest $F$ merged with an array of trees produced by the inner pipeline $Q$ evaluated on the \emph{joined} collection $F_j$.
Trees in $F_j$ are merged with a tree constructed over a sequence of variables $V$ before they are passed to $Q$.
The resulting array is stored at a non-empty path $p$ in output trees.
This can be written as:
\begin{eqnarray}
F \aggregate \lambda_{p}^{V}[F_j,Q] &=&
    \{ \tree \oplus \textsf{attach}(p, \textsf{array}(\lambda(V,F_j) \aggregate Q, \epsilon)
    )
    \mid \tree \in F \},
\\
\lambda(V,F_j) &=&
    \{ \tree \oplus \bigoplus_{q_i/d_i \in V} \json{ q_i: \textsf{eval}(d_i,\tree) } \mid \tree \in F_j \},
\end{eqnarray}
where $Q = s_1 \aggregate \dots \aggregate s_k$ is a MQuery with $k$ stages.
E.g., let $C_{\textsf{one}} = \{ \json{ \text{\_id}: 1 }\}$ and $C$ be a collection that consists of the documents
$\json{ \text{\_id}: 1, \text{a}: 3 }$ and
$\json{ \text{\_id}: 2, \text{a}: 5 }$.
Then, $\lambda^{b/5}_a[C, \mu_{\text{a}=\text{b}}]$ evaluated on input collection $C_{\textsf{one}}$ yields the document $\json{ \text{\_id}: 1, \text{a}: [\ \json{ \text{\_id}: 2, \text{a}: 5, \text{b}: 5 }\ ]\ }$.

The main difference to previous definition is that the array is constructed over a forest produced by an aggregation pipeline $q$ instead of over a single \emph{match} stage, and that additional nodes are added to input trees of $q$ that are interpreted as variable values.
The notion of variables is used in the aggregation framework within scoped expressions.
The convention is that variables are stored in trees with a reserved prefix (\emph{\$\$}) to avoid name clashes.
However, in the following, we will only project a reserved path $p_v$ into trees processed via the \emph{lookup} stage, and thus, can ignore this problem.

Finally, we characterize the \emph{graph-lookup} operation as a recursive function that is computed for each tree in an input forest $F$, and value definition in a non-empty sequence $d_1, \dots, d_k$.
Each recursion step selects trees of the joined collection $F_j$ that satisfy a condition $\varphi$, and that have the value $v$ stored at path $p_1$.
For the first step, $v$ is defined by one of the value definitions $d_1, \dots, d_k$, and, for each following step, the values at path $p_2$ of trees selected in the previous step is used.
Finally, an array of visited trees is constructed and stored at path $p$ in output trees.
This can be written as follows:
\begin{eqnarray}
F \aggregate \tau_{p, p_d, \varphi, D}^{p_1 = p_2}[F_j] &=&
    \bigcup_{\tree \in F, d_i \in D} \{
        \tree \oplus \textsf{attach}(p, \textsf{array}(\tau_{p_d, \varphi}^{p_1 = p_2}[F_j](v_i,1), \epsilon)) \},
\\
\tau_{p_d, \varphi}^{p_1=p_2}[F_j](v,n) &=&
        F_n \cup \bigcup_{\tree \in F_n} \tau_{p_d, \varphi}^{p_1=p_2}[F_j]( \textsf{eval}(p_2, \tree), n+1),
\end{eqnarray}
where
$F_n = F_j \aggregate \mu_{(p_1 = v) \wedge \varphi} \aggregate \rho_{\epsilon, p_d/n}$ denotes the forest produced in the $n$-th recursion step,
path $p_d$ stores the recursion depth, and
$v_i = \textsf{eval}(d_i, \tree)$ denotes the $i$-th starting value for the recursive search.
The recursion terminates at step number $m$ if $F_m = \emptyset$, i.e., if no record could be found that satisfies the match conditions.
In the aggregation framework, a maximum recursion depth can further be specified as a terminal condition which would be trivial to include in above formalism.
Additionally, the database performs cycle detection to avoid infinite recursion over circular references.

We further informally introduce new auxiliary notions in MQuery for the handling of complex terms and variables in operators of the language
(see App.~\ref{sec:appendix:semantics} for a formal definition).
A notion of variables and complex terms in MQuery is important because the structure of terms might not be fully known at compile-time, i.e., when the term in question contains a variable that was instantiated to another complex term through an earlier stage of the pipeline evaluation.
Thus, the flattened form of an instantatiated term must be constructed within MQuery.
To this end, we use new syntactic expressions in parameters of operators that evaluate to the flattened form of terms.
These may appear where value definitions can occur.
We use the following auxiliary notions over terms:
\begin{itemize}
\item $\textsf{fact}[p]$, which constructs a flattened term from a subtree of the input tree at path $p$. The subtree consists of a term in object form, i.e., with edges labeled by the index keys of the elements, and it is transformed into an array where each array element is an object of the form $\json{\text{k}: k_n, \text{v}: v_n }$, where $k_n$ is the index key of the element, and $v_n$ its value.
E.g., $\textsf{fact}[\epsilon]$ evaluated on the input tree $\json{ \text{0}: \text{"a"} }$ produces the array $[ \json{ \text{k}: \text{"0"}, \text{v}: \text{"a"} } ]$ representing the constant \emph{a}.

\item $\textsf{term}[\term, p, \varphi]$, which constructs a term in flattened form that is in instance of a segment of term $\term$.
The segment is determined by the filter condition $\varphi$ that must be satisfied by elements of the yielded flattened array.
E.g., $\varphi=(\exists \text{v})$ selects all elements where the field \emph{v} is defined, i.e., all indexed constants that appear in the term.
Each variable $v$ within the segment that has an instantiation in the node of the input tree at path $p.\textsf{name}(v)$ is replaced by the corresponding value in the array yielded.
The condition $\varphi$ can further be the constant \textsf{true}, in which case we write $\textsf{term}[\term, p] = \textsf{term}[\term, p, \textsf{true}]$.
E.g., $\textsf{term}[p(x), \epsilon]$ evaluated on the input tree
$\json{ \text{x}: 2 }$, where $\textsf{name}(x) = \text{x}$, yields the array
$[ \json{ \text{k}: \text{"0"}, \text{v}: \text{"p"} }, \json{ \text{k}: \text{"1"}, \text{v}: 2 } ]$.

\item $\textsf{subterm}[p,k]$, which yields a flattened subterm of another flattened term at a non-empty path $p$ in input trees.
The subterm consists of exactly the elements whose index key is prefixed by $k$ in the input tree.
Within the subterm the index key prefix is deleted in all elements.
E.g., $\textsf{subterm}[\text{a},1]$ evaluated on the input tree
$\json{ \text{a}: [
	\json{ \text{k}: \text{"0"},   \text{v}: \text{"p"} },
	\json{ \text{k}: \text{"1.0"}, \text{v}: \text{"q"} },
	\json{ \text{k}: \text{"1.1"}, \text{v}: 3 }
] }$
yields the array 
$[
	\json{ \text{k}: \text{"0"}, \text{v}: \text{"q"} },
	\json{ \text{k}: \text{"1"}, \text{v}: 3 }
]$
representing the subterm $q(3)$.

\item $p_1 \dashv p_2$, which constructs a flattened term that is an instance of the term at path $p_1$, where variables have been replaced with values of the term at path $p_2$.
A variable with an index key $k$ in the first term is replaced exactly by the values in the second term whose index key is prefixed by $k$.
E.g., $\text{a} \dashv \text{b}$ evaluated on
an input tree with the value
$[ \json{ \text{k}: \text{"0"}, \text{v}: \text{"p"} },
   \json{ \text{k}: \text{"1"}, \text{n}: \text{"x"} } ]$
at key \emph{a}, and
$[ \json{ \text{k}: \text{"0"}, \text{v}: \text{"p"} },
   \json{ \text{k}: \text{"1"}, \text{v}: \text{"q"} } ]$
at key \emph{b}
yields an array representing the term $p(q)$ where the variable \emph{"x"} in \emph{a} has been instantiated:
$[
	\json{ \text{k}: \text{"0"}, \text{v}: \text{"p"} },
	\json{ \text{k}: \text{"1"}, \text{v}: \text{"q"} }
]$.
\end{itemize}

\section{The Mongolog Language}
\label{sec:mongolog}

In this section, we define the Mongolog language.
First, we characterize it syntactically as a subset of the Prolog ISO standard.
We then define semantics of Mongolog programs in the formal framework of the MQuery language which suggests an implementation in the aggregation framework of MongoDB.
We provide additional examples of Mongolog programs and their MQuery translation in App. \ref{sec:appendix:translations}.

\subsection{The Syntax of Mongolog Programs}
\label{sec:syntax}

The syntax of Mongolog programs is based on the Prolog ISO standard~\cite{ISOProlog}
%
which
includes a huge set of built-in predicates from which we only consider a sub-set.
However, many predicates can be seen as syntactic sugar, and be defined through other predicates.
In this work, we will consider the following built-in predicates:
\begin{itemize}
    \item $\textsf{true}$ and $\textsf{false}$, which are control predicates that always succeed and fail respectively.
    
    \item $\textsf{ground}(\term)$, $\textsf{var}(\term)$ and $\textsf{nonvar}(\term)$, which are used to verify the type of term $\term$.
    These predicates are useful, e.g., to declare clauses of a rule based on the instantiation of an argument at evaluation time.
    Moreover, clauses can be pruned if they consist of a type-checking predicate, and it is known at compile-time that the term whose type is checked can not satisfy the predicate.
    
    \item $\term_i \doteq \term_j$, which reads as \emph{$\term_i$ unifies with $\term_j$}.
    The arguments $\term_i$ and $\term_j$ of the unification operator are terms that may contain variables.
    The unification problem is concerned with how variables in both terms can be substituted such that both terms are (syntactically) equal.
    To solve this problem, a substitution $\theta = \{ v_1 \mapsto \term_1, \dots, v_n \mapsto \term_n \}$ of distinct variables $v_1 \dots v_n$ in $\term_i,\term_j$ must be computed that makes $\term_i,\term_j$ syntactically equal if applied to both terms.
    This can be written in postfix notation as $\term_i\theta \equiv \term_j\theta$ where $\equiv$ denotes syntactic equality.
    Then, $\theta$ is called a \emph{unifier} of $\term_i$ and $\term_j$.
    In the following, we assume that two variables are syntactically equal only if they are instantiated to equal values, or if they have the same name (variable aliasing is not handled).
    
    
    \item $\textsf{limit}(\goal, k)$ and $\textsf{once}(\goal)$, which control the evaluation of a goal $\goal$ that may succeed multiple times by limiting the maximum number of solutions yielded.
    The unary \emph{once} predicate commits to the \emph{first} solution of its higher-order argument, and discards all other options, while the binary
    \emph{limit} predicate commits the the \emph{$k$-first} solutions.
    
    \item $\textsf{ignore}(\goal)$, which succeeds for the first solution of goal $\goal$, but also succeeds if no such solution exists.
    This is useful, e.g., for reading optional properties from a semi-structured database.
    
     \item $\plneg g$, which reads as \emph{not $g$}.
     The inference rule used for negation in Prolog is called \emph{negation as failure}.
     As the name suggests, the negation of a goal $g$ is assumed to hold in case no substitution of variables in $g$ can be found such that $g$ holds (closed world assumption).
    
    
    
    \item $\goal_i \plor \goal_j$, which reads as \emph{$\goal_i$ or $\goal_j$}, and is usually written in infix notation.
    Intuitively, the \emph{or} predicate succeeds for each solution of one of the goals $\goal_i,\goal_j$.
    However, with the exception of being embedded into a solution limiting predicate, or the \emph{cut} operator appearing in $\goal_i$, which might has the consequence that some solutions of $\goal_i$ and $\goal_j$ are not yielded.
    Any solution of the first goal is yielded before the ones of the second goal, if any.
    
    \item $\goal_i \rightarrow \goal_j \plor \goal_k$, which is called \emph{if-then-else} predicate.
    The construct commits to the first solution of the condition goal $\goal_i$.
    If such a solution exists, then goal $\goal_j$ is evaluated, and otherwise goal $\goal_k$ is evaluated.
    A variant of this predicate, called \emph{if-then} predicate is denoted as $\goal_i \rightarrow \goal_j$.
    It acts as $\goal_i \rightarrow \goal_j \plor \textsf{false}$, and, thus, fails in case the condition has no solution.
    
\end{itemize}

Those predicates, where one or more of the arguments are goals, are called \emph{meta predicates}.
Their higher-order arguments can be constructed at runtime in Prolog programs.
This is, however, not the case in Mongolog programs as the syntactic structure must be \emph{compiled into} the program, i.e.,
the functor and arity of the higher-order argument cannot be instantiated within a Mongolog program.

A predicate in a Prolog program can be recursively defined through clauses that have the predicate as a sub-goal.
General recursion is a powerful mechanism, e.g., it is well established that any problem stated in an iterative program can equivalently be expressed in a program with recursion.
Several different classes of recursion can be identified, among them the class of \emph{linearly recursive} programs where the definition of a recursive predicate consists only of one recursive sub-goal.
If we further assume that the recursively defined predicate has only two free variables, then it can be defined via the transitive closure of the predicate~\cite{Jagadish87}.
In the following, we consider the \emph{recursion-free} fragment of the Prolog language with an additional built-in predicate written as $p^{+}(t_1,t_2)$ used to compute the transitive closure of a 2-ary predicate $p$, where each $t_i$ is either a constant or variable.
Note that a Mongolog program with linear recursion over a binary predicate can be transformed into a recursion-free program with transitive closure predicate.

\subsection{The Semantics of Mongolog Programs}
\label{sec:semantics}

Logic programming languages often have \emph{model theoretic} semantics, i.e., they are characterized in a purely logical framework, and thus inherit desired properties such as commutativity of disjunction and conjunction.
Commutativity is a desired property for subgoals in a database query because operations can be ordered arbitrary.
But Prolog programs are not purely declarative, and are rather defined by \emph{operational} semantics, i.e., how they are processed by SLD-resolution~\cite{Vieille87}.
SLD resolution is a technique based on \emph{proof by contradiction} which is used for deciding the satisfiability of a propositional formula, and it uses syntactic unification for \emph{on demand} instantiation of fomulae.

In the following, we define operational semantics for the Mongolog language oriented towards an implementation in the aggregation framework of MongoDB.
Accordingly, we define semantics of Mongolog within the formal framework of MQuery, and attempt to capture operational characteristics in form of aggregation pipelines.

Intuitively, semantics of a Mongolog program $P$ can be characterized as a mapping $\mathfrak{M}_P$ from rules and EDB facts to the set of IDB facts that can be derived from $P$.
But such a \emph{bottom-up} characterization is not possible for a general Prolog program due to its procedural reading.
We rather define the semantics of Mongolog programs wrt. a query $\leftarrow \goal_1, \dots, \goal_n$ given by the user as a mapping to the set of instantiations of the query obtained by the evaluation of an aggregation pipeline:
\begin{eqnarray}
\mathfrak{M}_P(\leftarrow \goal_1, \dots, \goal_n) &=&
    \{
        ( \iota(\goal_1,\tree), \dots,
          \iota(\goal_n,\tree) )
        \mid
        \tree \in C_{\textsf{one}} \aggregate \phi(\goal_1, \dots, \goal_n)
    \},
\\
\phi(\goal_1, \dots, \goal_n) &=& \phi(\goal_1) \aggregate \dots \aggregate \phi(\goal_n),
\end{eqnarray}
where $\phi$ translates goals into aggregation pipelines,
$C_{\textsf{one}}$ denotes a collection with exactly one empty tree, and
$\iota(\goal,\tree) = \textsf{unflatten}(\textsf{eval}(\textsf{term}[\goal, \epsilon], \tree))$ is an instance of $\goal$ 
constructed from a tree that was created by the evaluation of goal $\goal$ wrt. tree $\tree$.
The overall idea is that we start with an empty tree, successively add variable instantiations to it in different stages of the pipeline, and finally instantiate variables in the user goal based on data in the output tree.
Note that this set-theoretical definition of $\mathfrak{M}_P$ ignores the $\sigma$-labeling of trees, and, thus, ordering of solutions is undefined.
The \textsf{unflatten} function is formally defined in App.~\ref{sec:appendix:semantics}.

As an example, let us consider the query $\leftarrow g_1, g_2$ against a database with the $C_{\textsf{hasPart/2}}$ collection, where $g_1=\textsf{hasPart}(x,y)$, $g_2=\textsf{hasPart}(y,z)$, and $x,y,z$ are variables.
The evaluation of the corresponding pipeline yields a tree for each instantiation of the variables such that $y$ is a part of $x$, and $z$ is a part of $y$.
One of the trees is $G=\json{
	\text{\_id}: 1,
	\text{vars}: \json{
		x: \text{"fridge1"},
		y: \text{"door1"},
		z: \text{"handle1"}
	}}$.
It instantiates the query literals in the following way:
$\iota(g_1,G)=\textsf{hasPart}(\textit{fridge1},\textit{door1})$
and
$\iota(g_2,G)=\textsf{hasPart}(\textit{door1},\textit{handle1})$.
Finally, the set $\mathfrak{M}_P$ is constructed from different possible instantiations of goal literals, e.g., the tuple $(\iota(g_1,G),\iota(g_2,G))$ is element of the set $\mathfrak{M}_P(\leftarrow g_1, g_2)$.

The semantics of a Mongolog program wrt. a goal can now be defined through translations of different types of predicates into aggregation pipelines.
First, we define a few rather trivial built-in predicates as a single \emph{match} operation in MQuery.
The constants \textsf{true} and \textsf{false} can trivially be seen as a \emph{match} operation with only constants in the criterion:
$\phi(\textsf{true}) = \mu_{1 = 1}$ and
$\phi(\textsf{false}) = \mu_{1 = 0}$.
In the following, we write $\mu_{1 = 1}$ as $\mu_\top$, and $\mu_{1 = 0}$ as $\mu_\bot$.
Obviously, it also holds that $F \aggregate \mu_\top = F$, and $F \aggregate \mu_\bot = \emptyset$.
Comparison predicates can be defined similarly by referring to the flattened form of the arguments within the match criterion:
\begin{equation}
    \phi(\term_1 = \term_2) = \mu_{
        \textsf{term}[\term_1, p_v] =
        \textsf{term}[\term_2, p_v]},
\quad\mathrm{and}\quad
    \phi(\term_1 \neq \term_2) = \mu_{
        \textsf{term}[\term_1, p_v] \neq
        \textsf{term}[\term_2, p_v]},
\end{equation}
where $\term_1,\term_2$ are terms that are flattened and evaluated within the \emph{match} stage to obtain arrays that are compared with each other.
Note that any two variables are considered equal under this definition if they have the same name, or are instantiated to the same value.
In general Prolog, two variables are also seen as equal if one is an \emph{alias} of the other.

We classify terms into three categories: variables, non-variables and grounded terms.
To test whether a term is a variable, a Mongolog program may consist of a sub-goal $\textsf{var}(\term)$.
It only succeeds when $\term$ is a variable that has not been instantiated in the input tree.
This can be written as: $\phi(\textsf{var}(\term)) = \mu_{\neg \exists p_v.\textsf{name}(\term)}$, where $\term$ is a variable.
If $\term$ is not a variable, then the \textsf{var} predicate can be translated into $\mu_\bot$.
Similarly, it holds that $\phi(\textsf{nonvar}(\term)) = \mu_\top$ if $\term$ is not a variable.
Finally, a term $t$ with variables is grounded in the input tree if there is no undefined value in its flattened form:
\begin{eqnarray}
    &\phi(\textsf{ground}(\term)) =
    \mu_{\textsf{term}[\term, p_v, (\neg \exists \text{v})] = []},
\end{eqnarray}
where $\neg \exists \text{v}$ is used to select only the variables of the flattened term, i.e., the elements of the array where the \emph{v} field is undefined.
It further holds that $\phi(\textsf{ground}(\term)) = \mu_\top$ if $\term$ is a ground term.

A variable is (possibly) instantiated through the evaluation of a predicate that has the variable as an argument.
The value of the variable is then stored in a node of the subtree at path $p_v$.
In the following, we make use of an auxiliary projection function that inspects a set of terms to generate a projection sequence with an element for each variable in the set of terms.
The value of a variable is given by the \emph{sub-term} of a flattened term at path $p$, where only the elements whose index key is prefixed by the variable index key are included:
\begin{eqnarray}
    \rho_{\textsf{term}}(\{ \term_1, \dots, \term_n \}, p) &=&
        \rho_{p_v, p_v.v_1/\textsf{subterm}[p, k_1], \dots, p_v.v_m/\textsf{subterm}[p, k_m]},
\end{eqnarray}
where
$\bigcup_{1 \leq i \leq n} \textsf{vars}(\term_i) = \{ (v_1,k_1), \dots, (v_m,k_m) \}$
is the set of variables in $\{ \term_1, \dots, \term_n \}$,  $v_i$ is the name of the i-th variable, and $k_i$ is its index key.
Note that $\textsf{subterm}[p, k]$ is used to construct a new term from all elements of the flattened term at path $p$ where $k$ is a prefix of the index key.

The data stored in database collections corresponds to EDB predicates in Mongolog programs with matching argument instantiations.
If an EDB predicate $\term = p(\term_1,\dots,\term_n)$ appears in a goal with variable arguments, then the input forest is enriched with instantiations of these variables within the sub-tree at path $p_v$, and an output tree is created for each possible instantiation.
This can be characterized as a \emph{lookup} stage over the trees that have an instance of $\term$:
\begin{eqnarray}
\phi(\term) &=&
    \lambda^{p_v/p_v}_{p}[
        C_{\term},
        \mu_{
            \textsf{term}[\term, p_v, (\exists \text{v})]
            \subseteq
            \textsf{fact}[\epsilon]
        }
        \aggregate \rho_{\textsf{term}}(\{ \term \}, \epsilon)
    ]
    \aggregate \omega_p
    \aggregate \rho_{p_v/p.p_v},
\end{eqnarray}
where $C_\term$ is a collection consisting of ground instances of $\term$.
Note that the matching condition succeeds iff each constant in the term appears at the same position in the ground instance of $\term$ drawn from $C_\term$.

The lookup operator $\lambda$ has the useful characteristic that the evaluation of the inner pipeline is scoped, i.e., stages in the inner pipeline do not interact with the number of trees in the output forest of the lookup stage.
Scoping is crucial for built-in predicates in Mongolog that rely on yielding only a limited number of solutions of a sub-goal such as the \textsf{limit} predicate.
These predicates can be defined through translations into lookup stages where the last stage in the inner pipeline is a \emph{limit} stage:
\begin{eqnarray}
\phi(\textsf{limit}(\goal, k)) &=&
    \lambda^{p_v/p_v}_{p}[C_{\textsf{one}}, \phi(\goal) \aggregate \kappa_{k}]
    \aggregate \omega_p
    \aggregate \rho_{p_v/p.p_v},
\\
\phi(\textsf{ignore}(\goal)) &=&
    \lambda^{p_v/p_v}_{p}[C_{\textsf{one}}, \phi(\goal) \aggregate \kappa_{1}]
    \aggregate \omega^{*}_p
    \aggregate \rho_{p_v/((\exists p.p_v) ? p.p_v : p_v)},
\\
\phi(\plneg \goal) &=&
    \lambda^{p_v/p_v}_{p}[C_{\textsf{one}}, \phi(\goal) \aggregate \kappa_{1}]
    \aggregate \mu_{p=[]}.
\end{eqnarray}

The \textsf{once} predicate can be seen as a special case of the \textsf{limit} predicate: $\textsf{once}(\goal) = \textsf{limit}(\goal, 1)$.
Note that negation semantics is defined through the failure of finding solutions of goal $\goal$ which is indicated by an empty array at path $p$.
We also observe that scoping is only needed for the \textsf{limit} predicate in case the input forest has more than one tree, and that this is not the case for the other two predicates as they need to succeed in case there are no solutions.

A scoped pipeline within a lookup stage can also be used to characterize the \emph{if-then} and \emph{if-then-else} predicates.
The idea is to use the \emph{lookup} operator to obtain a solution for the condition goal if any, and to evaluate the \emph{then} and \emph{else} goals conditionally depending on whether such a solution exists,
i.e., based on whether the array produced by the \emph{lookup} stage is empty or not. This can be written as:
\begin{eqnarray}
\phi(\goal_i \rightarrow \goal_j) &=&
    \lambda_{\textsf{if}}(g_i)
    \aggregate \lambda_{\textsf{then}}(g_j)
    \aggregate \omega_{p_t}
    \aggregate \rho_{p_v/p_t.p_v},
\\
\phi(\goal_i \rightarrow \goal_j \plor \goal_k) &=&
    \lambda_{\textsf{if}}(g_i)
    \aggregate \lambda_{\textsf{then}}(g_j)
    \aggregate \lambda_{\textsf{else}}(g_k)
    \aggregate \omega^{*}_{p_t}
    \aggregate \omega^{*}_{p_e}
    \aggregate \rho_{p_v/((\exists p_t.p_v) ? p_t.p_v : p_e.p_v)},
\\
\lambda_{\textsf{if}}(g_i) &=&
    \lambda^{p_v/p_v}_{p_i}[
        C_{\textsf{one}},
        \phi(g_i)
        \aggregate \kappa_{1}
    ],
\\
\lambda_{\textsf{then}}(g_j) &=&
    \lambda^{p_v/p_v,p_i/p_i}_{p_t}[
        C_{\textsf{one}},
        \omega_{p_i}
        \aggregate \rho_{\epsilon, p_v/p_i.p_v}
        \aggregate \phi(g_j)
    ],
\\
\lambda_{\textsf{else}}(g_k) &=&
    \lambda^{p_v/p_v,p_i/p_i}_{p_e}[
        C_{\textsf{one}},
        \mu_{p_i=[]}
        \aggregate \phi(\goal_k)
    ],
\end{eqnarray}
where $\goal_i$ is the condition goal,
$\goal_j$ is the action goal that is evaluated if the condition holds,
$\goal_k$ is the goal that is evaluated otherwise, and
$p_i$, $p_t$, and $p_e$ are the paths where the resulting arrays are stored.

A scoping mechanism is also important to define semantics of the \emph{cut} operator to distinguish predicates that are \emph{transparent} to the cut from the ones that are \emph{opaque} for it.
However, several techniques are known that transform a program with a cut into an \emph{entailment-equivalent} program without it~\cite{Baaz13}.
This mechanism is called \emph{cut-elimination}.
Here, we avoid the complication introduced by the cut, and only define semantics for the \emph{cut-free} fragment of Mongolog.


Next, we define semantics of the \emph{unification} predicate $\doteq$ based on an equivalence relation on terms.
We only approximate the usual semantics by saying that two terms are equivalent if they can be made syntactically equal up to variable aliasing, i.e., each variable may only have one unique name.
The simplified semantics for unification can be defined as:
\begin{equation}
\begin{aligned}
\phi(\term_1 \doteq \term_2) = \
    &\rho_{p_v, p_1/\textsf{term}[t_1, p_v],
               p_2/\textsf{term}[t_2, p_v]} \aggregate
    \\
    &\rho_{p_v, p_1/(p_1 \dashv\ p_2),
                p_2/(p_2 \dashv\ p_1)}
    \aggregate \mu_{p_1 = p_2}
    \aggregate \rho_{\textsf{term}}(\{ t_1, t_2 \}, p_1),
\end{aligned}
\end{equation}
where the first projection is used to store the terms $t_1$ and $t_2$ in flattened form at the paths $p_1$ and $p_2$ respectively, and
the second projection instantiates variables in each flattened term to the values in the other term where the index key is prefixed by the variable index key.
Note that a unifier $\theta$ may substitute a variable with an infinite term for unification problems such as $x \doteq f(x)$. Mongolog would, however, see $\theta = \{ x \mapsto f(x) \}$ as a unifier, and is, thus, unsound wrt. unification.

Let us now consider the translation of the \emph{or} predicate into an aggregation pipeline.
It can be characterized as a set of lookup operations, where the output trees are labeled with the index of the corresponding sub-goal within the predicate to distinguish solutions generated by different sub-goals in the sort operator.
Sorting is important for predicates that have a procedural reading, i.e., where the ordering of sub-goals matters for intended semantics.
The translation can be written as:
\begin{eqnarray}
\phi(\goal_1 \plor \dots \plor \goal_n) &=&
    \lambda_1
    \aggregate \dots 
    \aggregate \lambda_n
    \aggregate \rho_{p_v,p/[p_1, \dots, p_n]}
    \aggregate \omega_p
    \aggregate \omega_p
    \aggregate \sigma_{+p.p_i}
    \aggregate \rho_{p_v/p.p_v},
\\
\lambda_k &=&
    \lambda^{p_v/p_v}_{p_k}[C_{\textsf{one}}, \phi(\goal_k) \aggregate    \rho_{p_v,p_i/k}]
\end{eqnarray}
where $p_i$ is a constant path to the sub-goal index.
Note that the expression $[p_1, \dots, p_n]$ resolves to a nested array, and, thus, the \emph{unwind} stage is evaluated twice.

The transitive closure of a binary relationship $q$ can be defined through the \emph{graph-lookup} operator in MQuery.
However, an additional lookup must be performed in case the first argument $x$ of the predicate is not instantiated before the transitive closure is computed.
Each output tree of the operation consist of an instantation of the second argument $y$ to one of the values that are part of the transitive closure.
This can be written as:
\begin{eqnarray}
\phi(q^{+}(x, y)) &=&
    \begin{cases}
    \begin{aligned}
    Q_1[C_{q/2}](x) \aggregate
    &Q^{+}_{p,p_x}[C_{q/2}] \aggregate Q_2(y), &
        \text{if $x$ is a variable;} \\
    &Q^{+}_{p,c_x}[C_{q/2}] \aggregate Q_2(y), &
        \text{if $x=c_x$ is a constant;}
    \end{aligned}
    \end{cases}
\\
Q_1[C](x) &=&
    \lambda_{p_1}^{p_v/p_v}[
        C,
        \mu_{\nexists p_x} \aggregate
        \rho_{p_v,p_x/\epsilon.\text{1}}
    ] \aggregate
    \omega^{*}_{p_1} \aggregate
    \rho_{p_v,p_x/(\exists p_1.p_x ? p_1.p_x : p_x)};
\end{eqnarray}
where $Q^{+}_{p,d}[C] =
    \tau_{p_d,\textsf{true},d}^{\epsilon.1 = \epsilon.2}[C] \aggregate \omega_p$
computes the transitive closure over a binary relation in collection $C$ with a starting value for the first argument of the predicate that is defined by $d$,
$Q_2(y) = (\mu_{p.\text{2} = c_y})$ if $y=c_y$ is a constant, and
$Q_2(y) = (\mu_{\nexists p_y \vee p.\text{2} = p_y} \aggregate
        \rho_{p_v,p_y/\epsilon.\text{2}})$ if $y$ is a variable.
The paths $p$ and $p_1$ are used to temporary store the result of lookup operations, and $p_x = p_v.\textsf{name}(x)$ and $p_y = p_v.\textsf{name}(y)$ are the paths where the value of the variables is stored in output documents.

Let $p(v_1, \dots, v_n)$ be a $n$-ary IDB predicate with $m$ clauses, where each $v_i$ is a variable, and the $j$-th clause is defined as: $p(\term_{j1}, \dots, \term_{jn}) \leftarrow b_j$, where each $\term_{ji}$ is a term, and not necessary a variable, and, therefore, each clause may provide some instances of the predicate arguments.
In the following, we interpret
Mongolog rules as arguments of the \emph{or} predicate:
\begin{eqnarray}
\phi(p(v_1, \dots, v_n)) &=&
    \phi(g_1 \plor \dots \plor g_m),
\end{eqnarray}
where $g_i = (v_1 \doteq \term_{i1}, \dots, v_n \doteq \term_{in}, b_i)$ for all $1 \leq i \leq m$.
Here, we abstract away from the notion of rules and directly encode them in aggregation pipelines.
An alternative is to identify a Mongolog rule with a \emph{database view} that has an associated aggregation pipeline.
However, some rules cannot be evaluated in a bottom-up fashion,
e.g., when a variable argument of the head literal also appears in a negative literal in the body.
Thus, the notion of database views cannot be employed for the general case of rules in Mongolog.

\section{Optimization Techniques}
\label{sec:optimization}

The translations provided in Section~\ref{sec:mongolog} do not necessary imply the most efficient implementation.
In this section, we investigate a few simple techniques that can be employed to obtain more efficient aggregation pipelines that provide the same set of solutions.
We limit our investigation here to techniques that are specific to the Mongolog formalism, and exclude techniques that can be applied to Prolog programs or aggregation pipelines in the general case.

\subsection{Document Size Reduction}
The performance of query evaluation in MongoDB is influenced by the size of documents that are passed from one stage to another.
The larger the documents are, the more memory is needed to fit them, and more time is needed for input-output operations.
It is, thus, beneficial to ensure that these documents store only \emph{necessary data}, i.e., data needed for the evaluation of subsequent pipeline stages, or to present an answer to the user.

The answer, in the case of Mongolog programs, consists of variable instatiations such that some goal given by the user holds.
Our convention is to store these instantiations in a sub-tree at path $p_v$.
This allows us to restrict projections easily to include only variable instantiations while removing any potential temporary fields.
However, the size of $p_v$ can be reduced.
E.g., to answer a user query, the path $p_v$ may consist only of instantiations of variables that appear in the query goals.

With this optimization we reduce the size of trees between the evaluation of two sub-goals in a user query.
A sub-goal may, however, translate into a complex pipeline through clauses that define it, and that introduce new variables within their scope.
The \emph{$p_v$-reduction} can similarly limit the projection to the variables in body literals visited so far, however, the projection must retain variable instantiations that are still necessary in the call context such as variables that appear in the user query.
To this end, we define a \emph{contextualized} translation function $\phi^{'}$ that additionally receives a set of variables $V$ that may not be deleted from trees in the generated aggregation pipeline.
E.g., for a conjunctive query with $n$ subgoals, we have:
\begin{eqnarray}
\phi^{'}((\goal_1,\dots,\goal_n),V) &=&
    \phi^{'}(\goal_1,V_{1}) \aggregate \rho(V_{1})
    \aggregate \dots \aggregate
    \phi^{'}(\goal_n,V_{n}) \aggregate \rho(V_{n}),
\\
\rho(\{ (n_1,k_1), \dots, (n_j,k_j) \}) &=&
    \rho_{p_v.n_1/p_v.n_1,\dots,p_v.n_j/p_v.n_j},
\end{eqnarray}
where $V_{k} = V \cup \bigcup_{1 \leq i \leq k} \textsf{vars}(\goal_i)$ is the set of variables up to goal $\goal_k$, and each $n_i$ is the name of a variable.
Translations of other predicates in Mongolog can be contextualized in a similar way to take into account which variable instantiations must be retained, and, thus, avoid the projection of unnecessary data.

\subsection{Predicate Elimination}

Our definitions follow the top-down evaluation paradigm where the evaluation starts with a goal given by the user, and only rules are evaluated that are needed to satisfy this query.
Thus, all predicates are eliminated from the program that are not necessary to answer the query.
However, more predicates can be eliminated when considering variable instantations given in the user query, and in clauses that bind variables in body literals.
The idea is that, given the instantiation of some variables, the program can be \emph{partially evaluated}~\cite{Leuschel98} wrt. the goal to yield the pipelines $\mu_{\bot}$ or $\mu_{\top}$ indicating that some sub-goal must fail or succeed respectively without any new variable instantiations.

In case two terms $t_i,t_j$ do not contain variables, the terms are equal and also unify if they are syntactically equal.
Furthermore, if the terms are not \emph{unifiable}, then they are not equal, and also the unification predicate cannot succeed.
This can be written as:
\begin{equation*}
\begin{aligned}
&\phi(\term_i \doteq \term_j) = \phi(\term_i = \term_j) = \mu_\top,
    &&\text{if $\term_i \equiv \term_j$};
\\
&\phi(\term_i \doteq \term_j) = \phi(\term_i = \term_j) = \mu_\bot,
    &&\text{if $\neg\exists\theta: \term_i\theta \equiv \term_j\theta$}.
\end{aligned}
\end{equation*}
Similar predicate elimination rules can be defined for control structures considered in this work:
\begin{equation*}
\begin{aligned}
&\begin{aligned}
&\phi(\plneg \goal) = \mu_\bot,
    &&\text{ if $\phi(\goal) = \mu_\top$;}
&&\phi(\plneg \goal) = \mu_\top,
    &&\text{ if $\phi(\goal) = \mu_\bot$;}
\\
&\phi(\goal_i \rightarrow \goal_j) = \phi(\goal_j),
    &&\text{ if $\phi(\goal_i) = \mu_\top$;}
&&\phi(\goal_i \rightarrow \goal_j) = \mu_\bot,
    &&\text{ if $\phi(\goal_i) = \mu_\bot$;}
\\
&\phi(\goal_i \rightarrow \goal_j \plor \goal_k) = \phi(\goal_j), \ \ \ \ \ \ \ \ \
    &&\text{ if $\phi(\goal_i) = \mu_\top$;}
&&\phi(\goal_i \rightarrow \goal_j \plor \goal_k) = \phi(\goal_k),
    &&\text{ if $\phi(\goal_i) = \mu_\bot$;}
\end{aligned}
\\
&\begin{aligned}
&\phi(\goal_1 \plor \dots \plor \goal_k \plor \dots \plor \goal_n) =
    \phi(\goal),
    &&\text{if $\phi(\goal_k) = \mu_\bot$ and $\goal = (\goal_1 \plor \dots \plor \goal_{k-1} \plor \goal_{k+1} \plor \dots \plor \goal_n)$;}
\\
&\phi(\goal_1 , \dots , \goal_k , \dots , \goal_n) =
    \phi(\goal),
    &&\text{if $\phi(\goal_k) = \mu_\top$ and $g = (\goal_1 , \dots , \goal_{k-1} , \goal_{k+1} , \dots , \goal_n)$;}
\\
&\phi(\goal_1, \dots, \goal_n) = \mu_\bot,
    &&\text{if $\exists \goal_i \in \{ \goal_1, \dots, \goal_n \}: \phi(\goal_i) = \mu_\bot$;}
\\
&\phi(\textsf{ignore}(\goal)) = \mu_{\top},
    &&\text{if $\phi(\goal) = \mu_\top$ or $\phi(\goal) = \mu_\bot$; and}
\\
&\phi(\textsf{limit}(\goal,k)) = \phi(\goal),
    &&\text{if $\phi(\goal) = \mu_{\top}$ or $\phi(\goal) = \mu_\bot$}.
\\
\end{aligned}
\end{aligned}
\end{equation*}

Built-in predicates that verify the type of a term can be eliminated in case the type can be verified without evaluation of a query.
This maybe the case if the term was instantiated through a syntactic expression, or if its type can be inferred from operational characteristics of the program.
E.g., a free variable that was not bound by a predicate evaluated earlier must still be a variable, while it must be grounded if bound by an EDB predicate because the EDB consists only of ground facts.

\subsection{Lookup Elimination}

Another form of optimization is the reshaping of aggregation pipelines for better performance.
MongoDB includes a pipeline optimizer with rules for, e.g., pipeline sequence optimization, and projection optimization to reduce the size of documents in the pipeline.
However, we can make further optimizations based on knowing that the collection $C_\textsf{one}$ consists only of a single empty tree.

The idea is that, in some cases, we can replace $C_{\textsf{one}}$ by a collection from which the inner pipeline of a \emph{lookup} stage draws its input documents, and to transform the pipeline accordingly.
This is only possible if the program exhibits a specific structure where the array created by the lookup operation is unnested to process each element individually afterwards:
\begin{equation}
    C_{\textsf{one}}
    \aggregate S_1
    \aggregate \lambda^{p_v/p_v}_{p}[C_{\lambda}, S_2]
    \aggregate \omega_p
    \aggregate \rho_{p_v/p.p_v}
    =
    C_{\lambda}
    \aggregate S_1
    \aggregate S_2,
\end{equation}
where $S_1$ and $S_2$ are potentially empty aggregation pipelines, and $S_1$ may only consist of \emph{match} and \emph{project} stages.
This optimization applies to EDB predicates, the \emph{if-then}, and the \emph{limit} predicate.

Two subsequent lookup operators with the same joined collection can further be merged into a single lookup operation in some cases.
E.g., if the inner pipelines of subsequent lookup operators consist only of a single match operator, then the conditions of the match operators can be combined through disjunction in a single lookup operation.
It is further worth mentioning that the group operator, which we do not consider in this work, can be used to express arbitrary joins within a single collection~\cite{BotoevaCCX18}, and, thus, can be used to replace the lookup operator in some cases.


\section{Conclusion and Future Work}
\label{sec:conclusion}

In this paper, we have investigated Prolog as a querying language for MongoDB databases, and have defined operational semantics for a fragment of Prolog, that we have coined as Mongolog, based on the evaluation of query pipelines in the aggregation framework of MongoDB.
The evaluation is done based on a translation of Mongolog programs into MongoDB aggregation queries.
To this end, we have extended the MQuery formalism with notions and operators that are essential for the definition of predicate semantics in Prolog programs.
We have shown that semantics for a substantial set of built-in predicates of the ISO Prolog standard can be defined within MQuery, and, thus, be implemented with a MongoDB database.
We are currently working on an implementation of the Mongolog language with several additional built-in predicates including built-ins for the analysis and construction of terms and lists.

During our investigation it occurred that many predicates can only be expressed via the lookup operator even if no database join needs to be performed. This is due to the scoped data processing of the lookup operator.
We suggest to isolate this feature within an additional pipeline operator that runs an aggregation pipeline for each input document.
Furthermore, in version 5 of the aggregation framework, a new pipeline stage \emph{unionWith} was added.
It might be useful to write more efficient queries that evaluate the \emph{or} predicate.
However, it can only handle cases in its current form where the or predicate is evaluated before any other predicate because there is no way to pass in variables grounded in an earlier subgoal of the query.

Even though the proposed semantics approximates the intended semantics of Prolog programs well, it is not equivalent in some cases due to missing mechanisms for creating aliases of variables, and implicitly instantiating them.
It would be interesting to incorporate these mechanisms into our formalism in future work.
We would further like to investigate more optimization methods, e.g., optimizations based on tracking whether a free variable must have been grounded in an earlier subgoal of a query to reduce queries to simpler ones that run more efficient.
This would also help investigating under which circumstances a database view can be created for Mongolog rules which in turn would allow us to cover a larger class of recursion as MongoDB can compute the transitive closure of relations defined in database views.
Another type of recursion could be covered through loops over fixed size data as it can be expressed by the unwind operator.



\bibliography{mongolog}


\appendix

\section{Examples of MongoDB Aggregation Queries}
\label{sec:appendix:mongodb}

MongoDB databases have a powerful querying interface called the aggregation framework.
Aggregation queries are represented as pipelines of stages that receive input documents from the previous stage (or the input collection) to transform and yield them to the next stage.
In the following, we provide simple examples for the operators used in this work.
In the following, we further consider a collection called \textsf{inventory} consisting of the following documents:
\begin{lstlisting}[language=json]
{ "_id" : 1, "sku" : "almonds", "instock" : 120 },
{ "_id" : 2, "sku" : "bread",   "instock" : 80 },
{ "_id" : 3, "sku" : "cashews", "instock" : 60 },
{ "_id" : 4, "sku" : "pecans",  "instock" : 80 }
\end{lstlisting}

\subsection{Match}
\label{sec:appendix:match}
The \emph{match} operator filters documents that do not match the specified conditions.

\begin{example}
The following MQuery yields documents in the \textsf{inventory} collection where the \emph{instock} key has the value $80$:
\begin{equation*}
    \textsf{inventory} \aggregate \mu_{\text{instock}=80}
\end{equation*}
The corresponding MongoDB query is:
\begin{lstlisting}[language=json]
db.inventory.aggregate([
    { $match: { $expr: { $eq: ["$instock", 80] } } }
])
\end{lstlisting}
The evaluation of the pipeline produces the following results:
\begin{lstlisting}[language=json]
{ "_id" : 2, "sku" : "bread",   "instock" : 80 },
{ "_id" : 4, "sku" : "pecans",  "instock" : 80 }
\end{lstlisting}
\end{example}

\begin{example}
The following MQuery yields documents in the \textsf{inventory} collection where the \emph{instock} key has the value $80$:
\begin{equation*}
    \textsf{inventory} \aggregate \mu_{
        \textsf{term}[\term_1, p_v] =
        \textsf{term}[\term_2, p_v]}
\end{equation*}
\end{example}

\subsection{Unwind}
\label{sec:appendix:unwind}
The \emph{unwind} operator deconstructs an array field in input documents, and outputs a document for each element in the array.
For the example, let us consider a collection named \textsf{events} with the following document:
\begin{lstlisting}[language=json]
{ "_id" : 1, "tags" : ["work", "sports"] }
\end{lstlisting}

\begin{example}
The following MQuery yields a document for each tag of events in the \textsf{events} collection:
\begin{equation*}
    \textsf{events} \aggregate \omega_{\textsf{tags}}
\end{equation*}
The corresponding MongoDB query is:
\begin{lstlisting}[language=json]
db.events.aggregate([
    { $unwind: "tags" }
])
\end{lstlisting}
The evaluation of the pipeline produces the following results:
\begin{lstlisting}[language=json]
{ "_id" : 1, "tags" : "work" },
{ "_id" : 1, "tags" : "sports" }
\end{lstlisting}
\end{example}

\subsection{Project}
\label{sec:appendix:project}
The \emph{project} operator processes incoming documents by deleting, adding, and manipulating fields.

\begin{example}
The following MQuery adds a new Boolean field \emph{available} based on the number of items in stock in the \textsf{inventory} collection, and removes the number of items in stock:
\begin{equation*}
    \textsf{inventory} \aggregate \rho_{\text{sku},\text{available}/(\text{instock} > 0)}
\end{equation*}
The corresponding MongoDB query is:
\begin{lstlisting}[language=json]
db.inventory.aggregate([
    { $project: {
        "sku": 1,
        "available": { $gt: ["$instock", 0] }
    } }
])
\end{lstlisting}
The evaluation of the pipeline produces the following results:
\begin{lstlisting}[language=json]
{ "_id" : 1, "sku" : "almonds", "available": true },
{ "_id" : 2, "sku" : "bread",   "available": true },
{ "_id" : 3, "sku" : "cashews", "available": true },
{ "_id" : 4, "sku" : "pecans",  "available": true }
\end{lstlisting}
\end{example}

\subsection{Sort}
\label{sec:appendix:sort}
The \emph{sort} operator produces output documents in sorted order.

\begin{example}
The following MQuery yields documents in the \textsf{inventory} collection sorted by the number of items in stock, and by the \emph{sku} value in ascending order:
\begin{equation*}
    \textsf{inventory} \aggregate \sigma_{+\text{instock},+\text{sku}}
\end{equation*}
The corresponding MongoDB query is:
\begin{lstlisting}[language=json]
db.inventory.aggregate([
    { $sort: { "instock": 1, "sku": 1 } }
])
\end{lstlisting}
The evaluation of the pipeline produces the following results:
\begin{lstlisting}[language=json]
{ "_id" : 3, "sku" : "cashews", "instock" : 60 },
{ "_id" : 2, "sku" : "bread",   "instock" : 80 },
{ "_id" : 4, "sku" : "pecans",  "instock" : 80 },
{ "_id" : 1, "sku" : "almonds", "instock" : 120 }
\end{lstlisting}
\end{example}

\subsection{Limit}
\label{sec:appendix:limit}
The \emph{limit} operator limits the number of documents passed to the next pipeline stage.

\begin{example}
The following query limits the number of output documents to one:
\begin{equation*}
    \textsf{inventory} \aggregate \kappa_1
\end{equation*}
The corresponding MongoDB query is:
\begin{lstlisting}[language=json]
db.inventory.aggregate([
    { $limit: 1 }
])
\end{lstlisting}
The evaluation of the pipeline produces a single result, however, any of the documents in the \textsf{inventory} collection could be returned because the input collection of the \emph{limit} operator has \emph{undefined order}.
\end{example}

\begin{example}
The following query limits the number of output documents to one, and yields the document with the lowest id each time it is evaluated:
\begin{equation*}
    \textsf{inventory} \aggregate
    \sigma_{+\text{\_id}} \aggregate
    \kappa_1
\end{equation*}
The corresponding MongoDB query is:
\begin{lstlisting}[language=json]
db.inventory.aggregate([
    { $sort: { "_id": 1 } },
    { $limit: 1 }
])
\end{lstlisting}
The evaluation of the pipeline produces the following result:
\begin{lstlisting}[language=json]
{ "_id" : 1, "sku" : "almonds", "instock" : 120 }
\end{lstlisting}
\end{example}

\subsection{Lookup}
\label{sec:appendix:lookup}
The \emph{lookup} operator performs a left outer join to merge input trees with documents from the \emph{joined} collection.
For the following examples, we use an additional collection called \textsf{orders} consisting of the following documents:
\begin{lstlisting}[language=json]
{ "_id" : 1, "item" : "almonds", "quantity" : 2 },
{ "_id" : 2, "item" : "pecans", "quantity" : 1 }
\end{lstlisting}

\begin{example}
The following query joins information of the \textsf{inventory} and \textsf{orders} collection where their \emph{sku} and \emph{item} fields coincide:
\begin{equation*}
    \textsf{orders} \aggregate
    \lambda_{\text{a}}^{v/\text{item}}[
        \textsf{inventory},
        \mu_{\text{sku} = v}
    ]
\end{equation*}
The corresponding MongoDB query is:
\begin{lstlisting}[language=json]
db.orders.aggregate([
    { $lookup: {
        from: "inventory",
        let: { v: "$item" },
        pipeline: [
            { $match: { $expr: { $eq: [ "$$v", "$sku" ] } } }
        ]
        as: "a"
    } }
])
\end{lstlisting}
Note that variables passed into the \emph{lookup} operator are accessed with a different prefix then the fields of the input document.
The evaluation of the pipeline produces the following results:
\begin{lstlisting}[language=json]
{ "_id" : 1, "item" : "almonds", "quantity" : 2, a: [
    { "_id" : 1, "sku" : "almonds", "instock" : 120 } ] },
{ "_id" : 2, "item" : "pecans", "quantity" : 1, a: [
    { "_id" : 4, "sku" : "pecans",  "instock" : 80 }
\end{lstlisting}
\end{example}

\subsection{Graph Lookup}
\label{sec:appendix:graphlookup}
The \emph{graph-lookup} operator performs a recursive search on a collection for each input document, links the documents via two different fields during the search, and yields an output document for each terminal recursion path.
For the following example, we use an additional collection called \textsf{ancestors} consisting of the following documents:
\begin{lstlisting}[language=json]
{ "_id" : 1, "child": "a", "parent": "b" },
{ "_id" : 2, "child": "b", "parent": "c" },
{ "_id" : 3, "child": "b", "parent": "d" }
\end{lstlisting}

\begin{example}
Following MQuery computes the transitive closure of the ancestor relationship in collection \textsf{ancestors} for each value of the \emph{child} field:
\begin{equation*}
    \textsf{ancestors} \aggregate
    \tau^{\text{child}=\text{parent}}_{
        \epsilon.\text{a},
        \epsilon.\text{child}}[\textsf{ancestors}]
\end{equation*}
The corresponding MongoDB query is:
\begin{lstlisting}[language=json]
db.ancestors.aggregate([
    { $graphLookup: {
        from: "ancestors",
        startWith: "$child",
        connectToField: "child",
        connectFromField: "parent",
        as: "a"
    } } ])
\end{lstlisting}
The evaluation of the pipeline produces the following results:
\begin{lstlisting}[language=json]
{   "_id" : 1,
    "child" : "a",
    "parent" : "b",
    "a" : [ 
        {   "_id" : 1, "child" : "a", "parent" : "b" },
        {   "_id" : 2, "child" : "b", "parent" : "c" },
        {   "_id" : 3, "child" : "b", "parent" : "d" }
    ]
},
{   "_id" : 2,
    "child" : "b",
    "parent" : "c",
    "a" : [
        {   "_id" : 2, "child" : "b", "parent" : "c" },
        {   "_id" : 3, "child" : "b", "parent" : "d" }
    ]
},
{   "_id" : 3,
    "child" : "b",
    "parent" : "d",
    "a" : [
        {   "_id" : 2, "child" : "b", "parent" : "c" },
        {   "_id" : 3, "child" : "b", "parent" : "d" }
    ]
}
\end{lstlisting}
\end{example}

\section{Examples of Mongolog Queries}
\label{sec:appendix:translations}

Mongolog queries are formulated wrt. a program consisting of facts and rules, and translated into MQuery.
The following examples consist of Mongolog queries and programs, and their corresponding MQuery pipeline together with the evaluation results.
In the following, we denote the name of a variable $x$ as $p_x$, and assume it is the same name string as used in the formalization, e.g., $p_x = \textsf{name}(x) = \text{"x"}$.
Furthermore, $p_v = \text{"vars"}$ is the path where instantiations of variables are stored.

\subsection{EDB and IDB Predicates}
The basic building blocks of Mongolog programs are EDB and IDB predicates used to represent facts and rules of the program respectively.
In the following, we list example queries using EDB and IDB predicates.

\begin{example} Performing a conjunctive query over EDB predicates:
\begin{eqnarray*}
\textsf{hasPart}(\text{fridge1}, \text{door1}) &\leftarrow&
\\
\textsf{hasPart}(\text{door1}, \text{handle1}) &\leftarrow&
\\
\textsf{hasPart}(\text{door1}, \text{handle2}) &\leftarrow&
\\
&\leftarrow& \textsf{hasPart}(x,y), \textsf{hasPart}(y,z)
\end{eqnarray*}
The corresponding MQuery pipeline is:
\begin{gather*}
    C_{\textsf{one}} \aggregate
    \lambda_p^{p_v/p_v}[C_{\textsf{hasPart/2}} \aggregate Q_1] \aggregate
    \omega_p \aggregate
    \rho_{p_v/p.p_v} \aggregate
    \lambda_p^{p_v/p_v}[C_{\textsf{hasPart/2}} \aggregate Q_2] \aggregate
    \omega_p \aggregate
    \rho_{p_v/p.p_v}
    \\
    Q_1 = \mu_{
            \textsf{term}[
                \textsf{hasPart}(x,y), p_v,  (\exists \text{v})
            ] \subseteq \textsf{fact}[\epsilon]
        } \aggregate
        \rho_{p_v, p_v.\text{x}/\textsf{subterm}[\epsilon,\text{1}],
                   p_v.\text{y}/\textsf{subterm}[\epsilon,\text{2}]}
    \\
    Q_2 = \mu_{
            \textsf{term}[
                \textsf{hasPart}(y,z), p_v,  (\exists \text{v})
            ] \subseteq \textsf{fact}[\epsilon]
        } \aggregate
        \rho_{p_v, p_v.\text{y}/\textsf{subterm}[\epsilon,\text{1}],
                   p_v.\text{z}/\textsf{subterm}[\epsilon,\text{2}]}
\end{gather*}
The pipeline can be optimized via $\lambda$-elimination to:
\begin{equation*}
    C_{\textsf{hasPart/2}} \aggregate Q_1 \aggregate
    \lambda_p^{p_v/p_v}[C_{\textsf{hasPart/2}} \aggregate Q_2] \aggregate
    \omega_p \aggregate
    \rho_{p_v/p.p_v}
\end{equation*}
Note that the \emph{match} operators can further be simplified knowing that only $y$ has received an instantiation in the second sub-goal.
Further note that the second \emph{lookup} stage can also be eliminated through the use of the \emph{group} operator.
The evaluation of the pipeline produces the following results:
\begin{lstlisting}[language=json]
{ "_id": 1, "vars": { "x": "fridge1", "y": "door1", "z": "handle1" } },
{ "_id": 1, "vars": { "x": "fridge1", "y": "door1", "z": "handle2" } }
\end{lstlisting}
\end{example}

\begin{example} Defining and querying a reflexive relationship over an EDB predicate:
\begin{eqnarray*}
\textsf{hasPart}(\text{fridge1}, \text{door1}) &\leftarrow&
\\
\textsf{hasPart}_{\textsf{reflexive}}(x,y) &\leftarrow&
    \textsf{hasPart}(x,y) \plor \textsf{hasPart}(y,x)
\\
&\leftarrow& \textsf{hasPart}_{\textsf{reflexive}}(x,y)
\end{eqnarray*}
The corresponding MQuery pipeline is:
\begin{gather*}
\begin{aligned}
    C_{\textsf{one}} \aggregate
    &\lambda_{p_1}^{p_v/p_v}[
        C_{\textsf{one}} \aggregate
        \lambda_p^{p_v/p_v}[C_{\textsf{hasPart/2}} \aggregate Q_1] \aggregate
        \omega_p \aggregate \rho_{p_v/p.p_v} \aggregate
        \rho_{p_v,p_i/1}
    ] \aggregate
    \\
    &\lambda_{p_2}^{p_v/p_v}[
        C_{\textsf{one}} \aggregate
        \lambda_p^{p_v/p_v}[C_{\textsf{hasPart/2}} \aggregate Q_2] \aggregate
        \omega_p \aggregate \rho_{p_v/p.p_v} \aggregate
        \rho_{p_v,p_i/2}
    ] \aggregate
    \\
    & \rho_{p_v,p/[p_1,p_2]} \aggregate
    \omega_p \aggregate
    \omega_p \aggregate
    \sigma_{+p.p_i} \aggregate
    \rho_{p_v/p.p_v}
\end{aligned}
    \\
    Q_1 = \mu_{
            \textsf{term}[
                \textsf{hasPart}(x,y), p_v,  (\exists \text{v})
            ] \subseteq \textsf{fact}[\epsilon]
        } \aggregate
        \rho_{p_v, p_v.\text{x}/\textsf{subterm}[\epsilon,\text{1}],
                   p_v.\text{y}/\textsf{subterm}[\epsilon,\text{2}]}
    \\
    Q_2 = \mu_{
            \textsf{term}[
                \textsf{hasPart}(y,x), p_v,  (\exists \text{v})
            ] \subseteq \textsf{fact}[\epsilon]
        } \aggregate
        \rho_{p_v, p_v.\text{x}/\textsf{subterm}[\epsilon,\text{2}],
                   p_v.\text{y}/\textsf{subterm}[\epsilon,\text{1}]}
\end{gather*}
The evaluation of the pipeline produces the following results:
\begin{lstlisting}[language=json]
{ "_id": 1, "vars": { "x": "fridge1", "y": "door1" } },
{ "_id": 1, "vars": { "x": "door1", "y": "fridge1" } }
\end{lstlisting}
\end{example}

\begin{example} Defining and querying a transitive relationship over an EDB predicate:
\begin{eqnarray*}
\textsf{hasPart}(\text{fridge1}, \text{door1}) &\leftarrow&
\\
\textsf{hasPart}(\text{door1}, \text{handle1}) &\leftarrow&
\\
\textsf{hasPart}_{\textsf{transitive}}(x,y) &\leftarrow& \textsf{transitive}(\textsf{hasPart}(x,y))
\\
&\leftarrow& \textsf{hasPart}_{\textsf{transitive}}(\text{fridge1},y)
\end{eqnarray*}
The corresponding MQuery pipeline is:
\begin{equation*}
    C_{\textsf{one}} \aggregate
    \tau^{\epsilon.\text{1}=\epsilon.\text{2}}_{
        p, \text{"fridge1"}}[C_\textsf{hasPart/2}] \aggregate
    \omega_p \aggregate
    \mu_{p.1 \neq \text{"fridge1"}} \aggregate
    \rho_{p_v, p_v.\text{y}/p.1}
\end{equation*}
The evaluation of the pipeline produces the following results:
\begin{lstlisting}[language=json]
{ "_id": 1, "vars": { "y": "door1" } },
{ "_id": 1, "vars": { "y": "handle1" } }
\end{lstlisting}
\end{example}

\subsection{Control Predicates}
Control predicates provide some form of control over the inference process.
In the following, we provide some prototypical examples of Mongolog programs with control structures.

\begin{example} Limiting the solutions of a goal:
\begin{eqnarray*}
\textsf{bird}(\textit{tweety}) &\leftarrow&
\\
\textsf{bird}(\textit{tux}) &\leftarrow&
\\
&\leftarrow& \textsf{limit}(\textsf{bird}(x),1)
\end{eqnarray*}
The corresponding MQuery pipeline is:
\begin{gather*}
    C_{\textsf{one}} \aggregate
    \lambda_p^{p_v/p_v}[
        C_{\textsf{one}} \aggregate
        \lambda_p^{p_v/p_v}[C_{\textsf{bird/1}} \aggregate Q]
        \aggregate \omega_p \aggregate \rho_{p_v/p.p_v}
        \aggregate \kappa_1
    ]
    \aggregate \omega_p \aggregate \rho_{p_v/p.p_v}
    \\
    Q = \mu_{
            \textsf{term}[
                \textsf{bird}(x), p_v,  (\exists \text{v})
            ] \subseteq \textsf{fact}[\epsilon]
        } \aggregate
        \rho_{p_v, p_v.\text{x}/\textsf{subterm}[\epsilon,\text{1}]}
\end{gather*}
The pipeline can be reduced via $\lambda$-elimination to:
\begin{gather*}
    C_{\textsf{bird/1}} \aggregate Q \aggregate \kappa_1
\end{gather*}
The evaluation of the pipeline produces one of the following results:
\begin{lstlisting}[language=json]
{ "_id": 1, "vars": { "x": "tweety" } }
\end{lstlisting}
\begin{lstlisting}[language=json]
{ "_id": 1, "vars": { "x": "tux" } }
\end{lstlisting}
Two different results are possible because EDB facts are yielded by the database in undefined order.
\end{example}

\begin{example} Reading optional properties with the \textsf{ignore} predicate:
\begin{eqnarray*}
\textsf{person}(\text{fred}) &\leftarrow&
\\
\textsf{person}(\text{maria}) &\leftarrow&
\\
\textsf{hasChild}(\text{maria},\text{fred}) &\leftarrow&
\\
&\leftarrow& \textsf{person}(x), \textsf{ignore}(\textsf{hasChild}(x,y))
\end{eqnarray*}
The corresponding MQuery pipeline is:
\begin{gather*}
\begin{aligned}
    C_{\textsf{one}} \aggregate
    &\lambda_p^{p_v/p_v}[C_{\textsf{person/1}} \aggregate Q_1]
    \aggregate \omega_p \aggregate \rho_{p_v/p.p_v}
    \aggregate
    \\
    &\lambda_p^{p_v/p_v}[
        C_{\textsf{one}} \aggregate
        \lambda_p^{p_v/p_v}[C_{\textsf{hasChild/2}} \aggregate Q_2]
        \aggregate \omega_p \aggregate \rho_{p_v/p.p_v}
        \aggregate \kappa_1
    ]
    \aggregate \omega_p^{*} \aggregate \rho_{p_v/((\exists p.p_v)?p.p_v:p_v)}
\end{aligned}
    \\
    Q_1 = \mu_{
            \textsf{term}[
                \textsf{person}(x), p_v,  (\exists \text{v})
            ] \subseteq \textsf{fact}[\epsilon]
        } \aggregate
        \rho_{p_v, p_v.\text{x}/\textsf{subterm}[\epsilon,\text{1}]}
    \\
    Q_2 = \mu_{
            \textsf{term}[
                \textsf{hasChild}(x,y), p_v,  (\exists \text{v})
            ] \subseteq \textsf{fact}[\epsilon]
        } \aggregate
        \rho_{p_v, p_v.\text{x}/\textsf{subterm}[\epsilon,\text{1}],
                   p_v.\text{y}/\textsf{subterm}[\epsilon,\text{2}]}
\end{gather*}
The pipeline can be reduced via $\lambda$-elimination to:
\begin{gather*}
\begin{aligned}
    C_{\textsf{person/1}} \aggregate Q_1 \aggregate
    \lambda_p^{p_v/p_v}[
        C_{\textsf{hasChild/2}} \aggregate Q_2 \aggregate \kappa_1
    ]
    \aggregate \omega_p^{*} \aggregate \rho_{p_v/((\exists p.p_v)?p.p_v:p_v)}
\end{aligned}
\end{gather*}
The evaluation of the pipeline produces the following result:
\begin{lstlisting}[language=json]
{ "_id": 1, "vars": { "x": "fred" } },
{ "_id": 1, "vars": { "x": "maria", "y": "fred" } }
\end{lstlisting}
\end{example}

\begin{example} Non-monotonic reasoning via negation as failure:
\begin{eqnarray*}
\textsf{bird}(\textit{tweety}) &\leftarrow&
\\
\textsf{bird}(\textit{tux}) &\leftarrow&
\\
\textsf{penguin}(\textit{tux}) &\leftarrow&
\\
\textsf{canFly}(x) &\leftarrow& \textsf{bird}(x), \plneg \textsf{penguin}(x)
\\
&\leftarrow& \textsf{canFly}(x)
\end{eqnarray*}
The corresponding MQuery pipeline is:
\begin{gather*}
\begin{aligned}
    C_{\textsf{one}} \aggregate
    &\lambda_p^{p_v/p_v}[C_{\textsf{bird/1}} \aggregate Q_1]
    \aggregate \omega_p \aggregate \rho_{p_v/p.p_v}
    \aggregate
    \\
    &\lambda_p^{p_v/p_v}[
        C_{\textsf{one}} \aggregate
        \lambda_p^{p_v/p_v}[C_{\textsf{penguin/1}} \aggregate Q_2]
        \aggregate \omega_p \aggregate \rho_{p_v/p.p_v}
        \aggregate \kappa_1
    ]
    \aggregate \mu_{p=[]}
\end{aligned}
    \\
    Q_1 = \mu_{
            \textsf{term}[
                \textsf{bird}(x), p_v,  (\exists \text{v})
            ] \subseteq \textsf{fact}[\epsilon]
        } \aggregate
        \rho_{p_v, p_v.\text{x}/\textsf{subterm}[\epsilon,\text{1}]}
    \\
    Q_2 = \mu_{
            \textsf{term}[
                \textsf{penguin}(x), p_v,  (\exists \text{v})
            ] \subseteq \textsf{fact}[\epsilon]
        } \aggregate
        \rho_{p_v, p_v.\text{x}/\textsf{subterm}[\epsilon,\text{1}]}
\end{gather*}
The pipeline can be optimized via $\lambda$-elimination to:
\begin{gather*}
    C_{\textsf{bird/1}} \aggregate Q_1 \aggregate
    \lambda_p^{p_v/p_v}[
        C_{\textsf{penguin/1}} \aggregate Q_2 \aggregate \kappa_1
    ]
    \aggregate \mu_{p=[]}
\end{gather*}
The evaluation of the pipeline produces the following result:
\begin{lstlisting}[language=json]
{ "_id": 1, "vars": { "x": "tweety" } }
\end{lstlisting}
\end{example}

\subsection{Unification Predicate}
The unification predicate $\doteq$ attempts to make two terms syntactically equal through the instantiation of variables in both terms.

\begin{example} The unification of two constants:
\begin{eqnarray*}
&\leftarrow& 2 \doteq 2
\end{eqnarray*}
The corresponding MQuery is:
\begin{gather*}
    \rho_{p_v, p_1/\textsf{term}[2,p_v],
               p_2/\textsf{term}[2,p_v]} \aggregate
    \rho_{p_v, p_1/(p_1 \dashv\ p_2),
                p_2/(p_2 \dashv\ p_1)} \aggregate
    \mu_{p_1=p_2} \aggregate
    \rho_{p_v}
\end{gather*}
The evaluation of the pipeline on the $C_{\textsf{one}}$ collection produces the following result:
\begin{lstlisting}[language=json]
{ "_id": 1 }
\end{lstlisting}
\end{example}

\begin{example} The unification of a variable and a constant value instantiates the variable:
\begin{eqnarray*}
&\leftarrow& 2 \doteq x
\end{eqnarray*}
The corresponding MQuery is:
\begin{gather*}
    \rho_{p_v, p_1/\textsf{term}[2,p_v],
               p_2/\textsf{term}[x,p_v]} \aggregate
    \rho_{p_v, p_1/(p_1 \dashv\ p_2),
                p_2/(p_2 \dashv\ p_1)} \aggregate
    \mu_{p_1=p_2} \aggregate
    \rho_{p_v, p_v.\text{x}/\textsf{subterm}[p_2,\text{0}]}
\end{gather*}
The evaluation of the pipeline on the $C_{\textsf{one}}$ collection produces the following result:
\begin{lstlisting}[language=json]
{ "_id": 1, "vars": { "x": 2 } }
\end{lstlisting}
\end{example}

\begin{example} The unification of compound terms instantiates variables of each term to values given in the other term:
\begin{eqnarray*}
&\leftarrow& p(x,1) \doteq p(2,y)
\end{eqnarray*}
The corresponding MQuery is:
\begin{gather*}
    \rho_{p_v, p_1/\textsf{term}[p(x,1),p_v],
               p_2/\textsf{term}[p(2,y),p_v]} \aggregate
    \rho_{p_v, p_1/(p_1 \dashv\ p_2),
                p_2/(p_2 \dashv\ p_1)} \aggregate
    \mu_{p_1=p_2} \aggregate
    \\
    \rho_{p_v, p_v.\text{x}/\textsf{subterm}[p_2,\text{1}],
               p_v.\text{y}/\textsf{subterm}[p_1,\text{2}]}
\end{gather*}
The evaluation of the pipeline on the $C_{\textsf{one}}$ collection produces the following result:
\begin{lstlisting}[language=json]
{ "_id": 1, "vars": { "x": 2, "y": 1 } }
\end{lstlisting}
\end{example}

\begin{example} The unification of two free variables with different names:
\begin{eqnarray*}
&\leftarrow& p(x) \doteq p(y)
\end{eqnarray*}
The corresponding MQuery is:
\begin{gather*}
    \rho_{p_v, p_1/\textsf{term}[p(x),p_v],
               p_2/\textsf{term}[p(y),p_v]} \aggregate
    \rho_{p_v, p_1/(p_1 \dashv\ p_2),
               p_2/(p_2 \dashv\ p_1)} \aggregate
    \mu_{p_1=p_2} \aggregate
    \\
    \rho_{p_v, p_v.\text{x}/\textsf{subterm}[p_2,\text{1}],
               p_v.\text{y}/\textsf{subterm}[p_1,\text{1}]}
\end{gather*}
Note that no variable alias is created, and, thus, the match operation fails because the variables have different names.
\end{example}

\section{Semantics of Term Operations}
\label{sec:appendix:semantics}

Within the MQuery formalism, we represent a term $\term$ as an array of atomic subterms, where each subterm is represented as a tree encoding the index of the subterm, and its value, or its name if it is a free variable that has not been instantiated.
The instantiation of a variable is stored in trees processed by MQuery at a path $p_v.n_t$, where $p_v$ is a path, and $n_t = \textsf{name}(t)$ if $t$ is a variable.
The auxiliary operator $\textsf{flatten}$ flattens a term while instantiating its variables wrt. a tree $\tree$, where $\textsf{flatten}(\term,p_v,p_i,\tree)$ returns:
\begin{align*}
    &\{ \json{\text{k}: p_i, \text{v}: c_t} \},
        &&\text{if $t=c_t$ is a constant;}
    \\
    &\{ \json{\text{k}: p_i, \text{v}: \textsf{eval}(p_v.n_t,\tree)} \},
        &&\text{if $t$ is a variable and $\tree \models \exists p_v.n_t$;}
    \\
    &\{ \json{\text{k}: p_i, \text{n}: n_t} \},
        &&\text{if $t$ is a variable and $\tree \models \nexists p_v.n_t$;}
    \\
    &\{ \json{\text{k}: p_i.0, \text{v}: q} \} \cup
        \bigcup_{1 \leq k \leq n} \textsf{flatten}(t_k,p_v,p_i.k,\tree),
        &&\text{if $t=q(t_1,\dots,t_n)$.}
\end{align*}

An advantage of the array representation of terms is that we can map over elements of nested objects, e.g., to detect variables in them.
E.g., the \textsf{flatten} function evaluated on an empty tree can be used to
characterize the \textsf{vars} operator that selects all variables in a term, i.e., the ones that have a key \emph{"n"} storing their names instead of a key \emph{"v"} storing their value.
This can be written as:
\begin{equation}
\textsf{vars}(\term) =
        \{ (n_i,k_i) \mid
            \json{\text{k}: k_i, \text{n}: n_i} \in \textsf{flatten}(\term,\epsilon,\epsilon,\json{})
        \},
\end{equation}
where $n_i$ is the name of a variable in term $t$, and $k_i$ its index key.

More generally, each element of a flattened term is a tree $\tree_i$ that can be filtered based on whether it satisfies some criterion $\varphi$.
Let $p$ be a path where variable instantiations are stored.
We denote the array representation of a term $t$,
where every element satisfies a criterion $\varphi$ as $\textsf{term}[t,p,\varphi]$.
Given a tree $\tree$ with variable instantiations, we define the evaluation of $\textsf{term}[t,p,\varphi]$ as:
\begin{equation}
    \textsf{eval}(\textsf{term}[\term,p,\varphi],\tree) = \textsf{array}(
        \{ \tree_i \mid
            \tree_i \in \textsf{flatten}(\term,p,\epsilon,\tree)  \wedge
            \tree_i \models \varphi
        \}
    , \epsilon).
\end{equation}

Let $p$ be a path that stores a flattened term.
Each element of the flattened term has a unique index key
$k_i$ represented as a path.
A subterm of the flattened array with index key $k_p$ can be identified
by all elements of the flattened term whose index key is prefixed by $k_p$, i.e., where $k_i=k_p.k_p^{'}$ for some index key $k_p^{'}$.
We denote the subterm of a flattened term stored at path $p$ as $\textsf{subterm}[p,k_p]$, where $k_p$ is the index key of the subterm, and define its evaluation wrt. a tree $\tree$ as:
\begin{eqnarray}
    \textsf{eval}(\textsf{subterm}[p,k_p],\tree) &=&
    \begin{cases}
    \textsf{undefined}, &\text{if $\textsf{subterm}(p,k_p,\tree) = \emptyset$;}
    \\
    \textsf{array}(\textsf{subterm}(p,k_p,\tree),\epsilon) &\text{otherwise;}
    \end{cases}
    \\
    \textsf{subterm}(p, k_p, \tree) &=&
        \{  \json{  \text{k}: k_p^{'},
                    \text{v}: v_i,
                    \text{n}: n_i
            }
            \mid 
            \tree_i \in F \wedge 
            k_i = k_p.k_p^{'}
        \}
\end{eqnarray}
where
$k_i = \textsf{subtree}(\tree_i,\text{k})$,
$v_i = \textsf{subtree}(\tree_i,\text{v})$ if $\textsf{subtree}(\tree_i,\text{v}) \neq \textsf{null}$ and $v_i = \textsf{undefined}$ otherwise,
$n_i = \textsf{subtree}(\tree_i,\text{n})$ if $\textsf{subtree}(\tree_i,\text{n}) \neq \textsf{null}$ and $n_i = \textsf{undefined}$ otherwise, and
$F = \textsf{set}(\textsf{subtree}(\tree, p))$ is a set consisting of all elements of a term at path $p$ in tree $\tree$
with $\textsf{set}(v) = \{ v_1, \dots, v_n \}$ if $v=[v_1,\dots,v_n]$,
and $\textsf{set}(v) = \{ v \}$ otherwise.
Note that keys and paths are used interchangeably here.

Terms are presented to the user in usual Prolog syntax.
To this end, the auxiliary operator $\textsf{unflatten}$ constructs a term from its flattened form.
Let $x$ be a flattened n-ary term.
The evaluation of $\textsf{unflatten}(x)$ returns:
\begin{align*}
    &c_x,
        &&\text{if $x=\{ \json{\text{k}: 0, \text{v}: c_x} \}$;}
    \\
    &\textsf{var}(n_x),
        &&\text{if $x=\{ \json{\text{k}: 0, \text{n}: n_x} \}$;}
    \\
    &q(t_1,\dots,t_n),
        &&\text{if $\json{\text{k}: 0, \text{v}: q} \in x$ and $\forall 1 < i < n: t_i = \textsf{unflatten}(\textsf{subterm}(\epsilon,i,x))$,}
\end{align*}
where $\textsf{var}(n_x)$ yields the variable with the unique name $n_x$.

Let $p_1$ and $p_2$ be paths that store a flattened term.
We write $p_1 \dashv p_2$ to denote the flattened term that is created when variables in the term identified by $p_1$ are replaced with elements of the flattened term at path $p_2$ in some tree $\tree$.
The elements that instantiate a variable are exactly the ones whose index key is prefixed by the index key of the variable.
This can be written as:
\begin{equation}
\begin{aligned}
    \textsf{eval}(p_1 \dashv p_2, \tree) = \textsf{array}
        &(
        A_1 \cup 
        \{ \tree_2 \mid \tree_2 \in F_2 \wedge
            \\
            &(
            \exists p, \tree_1 \in V_1:
            \textsf{subtree}(\tree_1, \text{k}).p =
                \textsf{subtree}(\tree_2, \text{k})
        )\}
    , \epsilon),
\end{aligned}
\end{equation}
where $p$ is a possibly empty path, and
for all $1 \leq i \leq 2$ it holds that
$F_i = \textsf{set}(\textsf{eval}(p_i,\tree))$ is the set of elements in a flattened term at path $p_i$ in tree $\tree$,
$A_i = \{ \tree \mid \tree \in F_i \wedge \tree \models (\exists \text{v}) \}$ are its constant elements, and
$V_i = \{ \tree \mid \tree \in F_i \wedge \tree \models (\nexists \text{v}) \}$ are its variable elements.

Finally, facts that are stored in database collections are not represented as flattened arrays, and need to be converted into this form to be comparable to terms that appear in Mongolog programs.
Let $p$ be a path that stores a fact in some tree $\tree$ as a \emph{term document}, i.e., where the keys of the object are the index keys of atomic subterms of the fact, and the atoms their value.
We denote the flattened representation of the fact as $\textsf{fact}[p]$, and define its evaluation wrt. tree $\tree$ as:
\begin{equation}
    \textsf{eval}(\textsf{fact}[p],\tree) = \textsf{array}(\{
        \json{\text{k}: k_i, \text{v}: v_i}
        \mid 
        k_i \in K \wedge
        \textsf{subtree}(\tree,p.k_i) = v_i \neq \textsf{null}
    \}, \epsilon)
\end{equation}

\section{Default Ordering in MongoDB}
\label{sec:appendix:ordering}

Each value in MQuery is either of type literal, array, or object.
Literals are sorted according to natural order, i.e., either numerical or alphabetical.
Arrays are compared by their smallest elements if using the $<_{\mathcal{M}}$ operator.
Objects are treated as sequences of key-value pairs in MongoDB databases, and, for comparison, the keys and values of two objects are compared in sequence-order.
However, the key-value pairs in MQuery trees are not ordered, and, thus, another metric must be employed.
Instead, we first build naturally ordered sequences
$k_{x1} <_{\mathcal{N}} \dots <_{\mathcal{N}} k_{xn}$ and
$k_{y1} <_{\mathcal{N}} \dots <_{\mathcal{N}} k_{ym}$ of all keys of the objects $v_x$ and $v_y$ respectively, and then compare keys and values in sequence order.
Hence, for two typed values $v_x$ and $v_y$, we say it is always true that $v_x <_{\mathcal{M}} v_y$, while:
\begin{equation*}
\begin{aligned}
&v_x = \textsf{null} \wedge v_y \neq \textsf{null}; &&
\\
&v_x <_{\mathcal{N}} v_y,
    &&\text{if $v_x$ and $v_y$ are literals;}
\\
&\textsf{min}(v_x) <_{\mathcal{M}} \textsf{min}(v_y),
    &&\text{if $v_x$ and $v_y$ are arrays;}
\\
&
\begin{aligned}
\exists k_{xi}: \
        ( &\forall 0<j<i: \
           k_{xj} =_{\mathcal{N}} k_{yj} \wedge
            v_{xj} =_{\mathcal{M}} v_{yj}) \wedge
        \\
        (  &k_{xi} <_{\mathcal{N}} k_{yi} \vee (
            k_{xi} =_{\mathcal{N}} k_{yi} \wedge 
            (v_{xi} <_{\mathcal{M}} v_{yi})
        ))
\end{aligned}
    &&\text{if $v_x$ and $v_y$ are objects;}
\end{aligned}
\end{equation*}
where
$v_{ij} = \textsf{subtree}(v_i,k_{ij})$, and
$\textsf{min}([v_1,\dots,v_n]) = v_m$ with $v_m \in \{ v_1,\dots,v_n \}$, and for all $v_k \in \{ v_1,\dots,v_n \} \setminus \{ v_m \}$ it holds that $v_m <_{\mathcal{M}} v_k$.
In case $v_x$ and $v_y$ are of different type, a fixed order among data types is used by MongoDB databases.
That is, $v_x <_{\mathcal{M}} v_y$ also holds if $v_x$ is a literal, and $v_y$ is either an object or array, and if $v_x$ is an object, and $v_y$ is an array.
Further note that, in MongoDB databases, $>_{\mathcal{M}}$ compares the largest array elements instead.
This is a problematic choice as an array can be smaller \emph{and} larger then another one.

\end{document}